\documentclass[useAMS,usenatbib]{mn2e}

\usepackage{natbib}

\usepackage{comment}
\usepackage{latexsym,graphicx}
\usepackage{color}
\usepackage{fixltx2e}
\usepackage{verbatim} \usepackage{float}
\usepackage{amsmath,amssymb}
\usepackage{times}

\usepackage{setspace}
\usepackage{rotating}
\usepackage{pdflscape}
\usepackage{subfig}
\usepackage{xcolor}
\usepackage{ulem}

\usepackage[squaren, Gray, cdot]{SIunits}

\newcommand\hii{H\,{\sc ii} \,}

\definecolor{purple}{rgb}{0.5, 0.0, 0.5}

\def\apgt{\ {\raise-.5ex\hbox{$\buildrel>\over\sim$}}\ }
\def\aplt{\ {\raise-.5ex\hbox{$\buildrel<\over\sim$}}\ }

\usepackage{hyperref}

\let\oldhat\hat
\renewcommand{\hat}[1]{\oldhat{\mathbf{#1}}}

\renewcommand{\degree}{\ensuremath{^\circ}}

\title[3D MHD astrospheres]{3D MHD astrospheres: applications to \textcolor{black}{IRC-10414} 
and Betelgeuse}

\author[D. M.-A.~Meyer et al.]
       {D. M.-A.~Meyer\thanks{E-mail: dmameyer.astro@gmail.com}$^{1}$, A.~Mignone$^{2}$, 
       M.~Petrov$^{3}$, K.~Scherer$^{\textcolor{black}{4,5}}$, P. F.~Vel\'azquez$^{6}$, P.~Boumis$^{7}$ \\ 
       $^{1}$ Universit\" at Potsdam, Institut f\" ur Physik und Astronomie, 
       Karl-Liebknecht-Strasse 24/25, 14476 Potsdam, Germany \\  
       $^{2}$ Dipartimento di Fisica Generale Facolt\`a di Scienze M.F.N., 
       Universit\`a degli Studi di Torino, Via Pietro Giuria 1, I-10125 Torino, Italy  \\
       $^{3}$ Max Planck Computing and Data Facility (MPCDF), 
       Gießenbachstrasse 2, D-85748 Garching, Germany     \\       
       $^{4}$ Institut f\" ur Theoretische Physik, Lehrstuhl IV: Plasma-Astroteilchenphysik,
       Ruhr-Universit\" at Bochum, D-44780 Bochum, Germany   \\
       $^{5}$ Research Department, Plasmas with Complex Interactions, 
       Ruhr-Universit\" at Bochum, 44780 Bochum, Germany   \\
       $^{6}$ Instituto de Ciencias Nucleares, Universidad Nacional Aut\' onoma 
       de M\' exico, CP 04510. Mexico City, Mexico    \\ 
       $^{7}$ Institute for Astronomy, Astrophysics, Space Applications and Remote 
       Sensing, National Observatory of Athens, 15236, Penteli, Greece  \\        
       }

\begin{document}

\date{Received; accepted}

\maketitle

\label{firstpage}

\begin{abstract} 
A significative fraction of all massive stars in the Milky Way move supersonically through  
their local interstellar medium (ISM), producing bow shock nebulae by wind-ISM 
interaction. The stability of these observed astrospheres around cool massive stars 
challenges precedent two-dimensional (magneto-)hydrodynamical simulations of their 
surroundings. 
We present three-dimensional magneto-hydrodynamical (3D MHD) simulations of the 
circumstellar medium of runaway M-type red supergiant stars moving with velocity 
$v_{\star}=50\, \rm km\, \rm s^{-1}$. We treat the stellar wind with a Parker 
spiral and assume a $7\, \rm \mu G$ magnetisation of the ISM. Our free 
parameter is the angle $\theta_\mathrm{mag}$ between ISM flow and magnetisation, 
taken to $0\degree$, $45\degree$ and $90\degree$. 
It is found that simulation dimension, coordinate systems and grid effects 
can greatly affect the development of the modelled \textcolor{black}{astrospheres}. 
Nevertheless, as soon as the ISM flow and magnetisation directions differs by more 
than a few degrees ($\theta_\mathrm{mag}\ge5\degree$), the bow shock is stabilised, 
most clumpiness and ragged structures vanishing.  
The complex shape of the bow shocks induce important projection effects, 
e.g. at optical H$\alpha$ line, producing complex of astrospheric morphologies. 
We speculate that those effects are also at work around earlier-type massive  
stars, which would explain their diversity of their observed arc-like nebula 
around runaway OB stars.
Our 3D MHD models are fitting well observations of the astrospheres of several 
runaway red supergiant stars. The results interpret the smoothed astrosphere of 
IRC-10414 and Betelgeuse ($\alpha$Ori) are stabilised by an organised, 
non-parallel ambient magnetic field. Our findings suggest that IRC-10414 is 
currently in a steady \textcolor{black}{state} of its evolution, and that Betelgeuse's bar is of 
interstellar origin. 
\end{abstract}



\section{Introduction}
\label{sect:intro}

Massive stars ($M_{\star}\ge8\, \rm M_{\odot}$) are preponderant 
engines in the cycle of matter of the interstellar 
medium (ISM) of galaxies~\citep{maeder_2009}.
They blow strong stellar winds, and, by synthetizing heavy chemical 
elements in their interiors, massive stars typically evolve from a 
long, hot, main-sequence phase to a short, cold red supergiant phase~\citep{ekstroem_aa_537_2012}. 
They may also experience series of eruptive, luminous blue mass-losing events and/or 
finally finish their lives as hot Wolf-Rayet stars, respectively~\citep{vink_asp_353_2006}. 
The number, duration and stellar surface properties of these successive 
evolutionary phases are mostly determined by their initial mass, rotation 
rate~\citep{brott_aa_530_2011a,2020arXiv200408203S}, but also chemical 
composition~\citep{sander_mnras_491_2020}, which uniquely characterise the 
evolution and fate of massive stars~\citep{woosley_rvmp_74_2002}. 
Throughout their lives, they lose mass and radiate strong 
ionizing photons, which both shape their surroundings as circumstellar bubble 
nebulae made of stellar wind and ISM material~\citep{weaver_apj_218_1977}.

The internal structures of wind-blown bubbles reflect the past stellar evolution 
of massive stars~\citep{garciasegura_1996_aa_316ff,freyer_apj_638_2006,
freyer_apj_594_2003,dwarkadas_apj_667_2007,2010MNRAS.405.1047G, 
vanmarle_584_aa_2015,meyer_mnras_493_2020}. 
These circumstellar nebulae are punctual valves located in the ISM which release 
energy, momentum and heavy elements that considerably enrich their local ambient 
medium~\citep{langer_araa_50_2012}. 
Some massive stars ($\lesssim 40\, \rm M_{\odot}$) eventually die as 
core-collapse supernova, a final explosive event marking their 
evolution~\citep{woosley_araa_44_2006,smartt_araa_47_2009}. 
It engenders a propagating blastwave, first going through their circumstellar 
medium, before further expanding into the pristine 
ISM~\citep{Chevalier_araa_15_1977,weiler_araa_25_1988}. 
When the ejecta material of the defunct star shocks its surroundings, it 
produces nebulae of complex morphologies called supernova remnants, 
displaying unusual patterns of enriched gas emitting light throughout the 
whole electromagnetic spectrum, by means of both thermal and non-thermal 
emission~\citep{weiler_araa_25_1988}.

About $4$$-$$10\%$ of all massive main-sequence stars are moving supersonically 
though the ISM~\citep{renzo_aa_624_2019}.  
This happens when (i) the stars are ejected by gravitational swing from their 
parent stellar clusters~\citep{lada_araa_41_2003}, (ii) by the explosive dissociation 
of binary systems~\citep{blaauw_bain_15_1961,gies_apjs_64_1987,hoogerwerf_aa_365_2001,
dincel_mnras_448_2015} 
\textcolor{black}{
or (iii) when massive binary system captures a third star, 
producing an unstable triple binary system from which one of the original binary 
component is ejected~\citep{gvaramadze_mnras_410_2011}. 
}
Hence, massive stars can move with high space velocities 
(from tens to hundreds of $\rm km\, \rm s^{-1}$), and, therefore, the wind-blown 
bubble of a runaway massive star is distorted as a bow shock 
nebula~\citep{wilkin_459_apj_1996}. 
Such a circumstellar structures can mainly be observed in the context of hot 
massive runaway stars at various 
wavelengths~\citep{gull_apj_230_1979,kaper_apj_475_1997,brown_aa_439_2005,
deBecker_mnras_471_2017}. However, they are mostly visible in the infrared 
waveband~\citep{vanburen_apj_329_1988,vanburen_aj_110_1995,noriegacrespo_aj_113_1997,
povich_apj_689_2008,peri_aa_538_2012,peri_aa_578_2015,
kobulnicky_apjs_227_2016,kobulnicky_aj_154_2017} whose emission are governed by dust 
physics~\citep{henney_mnras_486_2019,henney_486_mnras_2019,
henney_2019_arXiv190400343H,2019arXiv190700122H}. 
Stellar wind bow shocks also exhibit polarized 
emission~\citep{shrestha_mnras_477_2018,shrestha_mnras_500_2021}
and they are suspected to be cosmic-ray 
accelerators~\citep{valle_ApJ_864_2018,benaglia_mnras_503_2021}. 
While massive stars run away, stellar evolution keeps going, and, 
consequently, a significant fraction of all core-collapse supernova remnants 
involve a runaway progenitor~\citep{eldridge_mnras_414_2011}. 
The wind bubble and stellar wind bow shocks forming around runaway stars, ending their 
lives as red a supergiant, constitute the pre-supernova 
circumstellar medium inside of which the blastwave will subsequently 
expand~\citep{vanmarle_aa_444_2005,gvaramadze_aa_454_2006,vanmarle_aa_478_2008,  
gvaramadze_aa_454_2006,chiotellis_aa_537_2012,vanmarle_aa_541_2012, 
broersen_mnras_441_2014, meyer_mnras_450_2015,chiotellis_mnras_502_2021, 
meyer_mnras_502_2021}.

Bow shocks form when the stellar wind of evolved, red 
supergiant massive stars interact with their ambient medium~\citep{meyer_2014bb}. 
%
Because of their \textcolor{black}{minimal} terminal wind velocities ($\sim 20\, \rm km\, \rm s^{-1}$), 
the bow shocks of red supergiant stars are much smaller than those forming around 
moving hot OB stars, and, because of the low effective temperature 
of the star, the physics taking place inside of them is also more 
complex~\citep{teyssier_aa_545_2012,matsuura_mnras_437_2014}. 
Their winds are the site of rich chemical reactions~\citep{cannon_mnras_502_2021}, 
and the  clumps aggregating therein lead to the formation of large 
molecules~\citep{ogorman_aa_573_2015,montarges_mnras_485_2019}. 
The dusty, neutral winds of red supergiant stars eventually interact with the hot, 
ionised gas of either the remaining \hii region of the OB star they descent from or of a 
neighbouring stellar cluster including, e.g. hot Wolf-Rayet 
star(s)~\citep{meyer_2014a,mackey_aa_586_2016}. 
The neutral-ionised interface generates a photoionised-confined shell separating 
molecular gas from its ionized surroundings, which can be 
observed~\citep{2014Natur.512..282M}. 
\textcolor{black}{
Note that similar features based of enhanced neutral walls also exist in the 
heliosheath and in the astrosheaths of the 
Sun~\citep{izmodenov_apj_594_2003,scherer_aa_563_2014}. 
}
Three known bow-shock-producing red supergiant stars have been observed so far, 
namely that of Betelgeuse~\citep{noriegacrespo_aj_114_1997}, 
$\mu$Cep~\citep{2012A&A...537A..35C} and IRC-10414~\citep{Gvaramadze_2013}, 
raising the question of their apparent smooth shape despite the instabilities 
predicted to be at work therein~\citep{dgani_apr_461_1996,dgani_apj_461_1996}. 
The active role of the external ionization and magnetic fields has been shown 
to be a potential stabiliser in several numerical 
simulations~\citep{meyer_2014a,vanmarle_584_aa_2015}.

Stellar wind bow shocks from intermediate-mass and massive stars have been studied 
in many previous works. Several numerical models were produced to understand 
the functioning and emission properties of bow shock surroundings 
hot~\citep{blondin_na_57_1998,comeron_aa_338_1998,meyer_2014bb,
meyer_obs_2016,meyer_mnras_464_2017,green_aa_625_2019,meyer_mnras_496_2020} 
and cold~\citep{brighenti_mnras_277_1995,
WareingZijlstraOBrien_2007_MNRAS_382_1233_AGB_bowshocks,wareing_apj_660_2007,vanmarle_apjl_734_2011,
villaver_apj_748_2012,cox_aa_537_2012,vanmarle_aa_561_2014,acreman_mnras_456_2016} 
runaway massive stars. 
The runaway red supergiant stars 
Betelgeuse~\citep{mackey_apjlett_751_2012,mohamed_aa_541_2012,vanmarle_aa_537_2012}
and IRC-10414~\citep{meyer_2014a} motivated 
dedicated studies tailored to their environment. 
Until recently, most simulations, except very few, were two-dimensional 
hydrodynamical models. Over the past few years, \textcolor{black}{several} three-dimensional 
hydrodynamical~\citep{blondin_na_57_1998,mohamed_aa_541_2012} 
and \textcolor{black}{three-dimensional (magneto-)hydrodynamical (3D MHD)} models have been  
performed~\citep{gvaramadze_mnras_474_2018,katushkina_MNRAS_473_2018,
scherer_mnras_493_2020,baalmann_aa_650_2021}. However, no one so far has 
investigated the appearance of bow shocks from moving red supergiant stars 
using 3D MHD simulations. 
\textcolor{black}{
Therefore, the question is, which differences exist between two- and 
three-dimensional simulations of bow shocks around cool stars?
}
How does a 3D MHD Eulerian bow shock model \textcolor{black}{compare to that of its} 2D hydrodynamical 
counterpart? Does it affect the optical emission properties of the \textcolor{black}{astrosphere}? 
Can one directly compare 3D MHD models to observed bow shocks and pronounce on 
the stability and the surroundings of, e.g. IRC-10414 or Betelgeuse? 
In this paper, we investigate \textcolor{black}{employing} 3D MHD numerical simulations the 
effects of an ISM magnetic field that is not aligned with the direction of 
motion of a red supergiant star onto the structure and emission properties  
of its associated stellar wind bow shock.

Our study is organised as follows. We introduce the reader to the numerical 
methods used to perform our 3D MHD simulations of stellar wind bow shocks around 
runaway red supergiant stars in Section~\ref{sect:method}. 
We describe the results and analyse therein the effects of the presence of 
the ISM magnetic field onto the organisation, stabilisation and emission properties
of the bow shock nebulae in Section~\ref{sect:results}. 
Our results are further discussed in Section~\ref{sect:discussion}, and 
we conclude in Section~\ref{sect:conclusion}.


\begin{table*}
	\centering
	\caption{
	List of models in our study. The columns inform on the simulation labels, 
	the angle $\theta_{\rm mag}$ (in degrees) between the direction of the 
	ISM magnetic field and the direction of stellar motion, the strength of the 
	ISM magnetic field $B_{\rm ISM}$ (in $\mu \rm G$), whether the star 
	moves along the axis $Oz$ of the Cartesian computational domain, the 
	name of the red supergiant star to which the model is tailored to, whether 
	optically-thin radiative cooling and heating is included, and, finally, 
	the general purpose of each runs. 
	}
	\begin{tabular}{lcccccr}
	\hline
	${\rm {Model}}$ & $\theta_{\rm mag} (\degree)$  & $B_{\rm ISM} (\mu \rm G)$ & $\vec{v_{\star}} 
	\parallel\, $ $Oz$  & Star & Heating and radiative cooling & Purpose \\ 
	\hline   
	Run-test-1                   &  $0$    & $0$ & yes  &    IRC-10414$^{a}$    &  
	no                               &  test grid effects   \\ 
	Run-test-2                   &  $0$    & $0$ & no   &    IRC-10414$^{a}$    &  
	no                               &  test grid effects   \\ 
	Run-$\theta_{\rm mag}$-0     &  $0$    & $7$ & no   &    IRC-10414$^{a}$    &  
	heating/cooling for ionized gas  & bow shock stability study     \\	
	Run-$\theta_{\rm mag}$-45    &  $45$   & $7$ & no   &    IRC-10414$^{a}$    &  
	heating/cooling for ionized gas  & bow shock stability study    \\
	Run-$\theta_{\rm mag}$-90    &  $90$   & $7$ & no   &    IRC-10414$^{a}$    &  
	heating/cooling for ionized gas  & bow shock stability study   \\
	Run-$\alpha$Ori              &  $90$   & $7$ & no   &    Betelgeuse$^{b}$   &  
	heating/cooling for ionized gas  & comparison with observations     \\	
	\hline    
	\end{tabular}
\label{tab:models}\\
\footnotesize{
\textcolor{black}{
Stellar wind properties are taken from (a)~\citet{meyer_2014a} 
; (b)~\citet{vanmarle_aa_561_2014}, respectively.
}
}
\end{table*}

\section{Method}
\label{sect:method}

This section presents both the governing equations that we solve to simulate 
the circumstellar medium of evolved, cool runaway massive stars, together with 
the initial conditions used in our study.

\subsection{Governing equations}
\label{sect:method_equation}

We consider the problem of a circumstellar nebulae generated by stellar wind-ISM 
interaction around an evolved massive stellar object supersonically 
moving through a magnetised ISM. 
The evolution of the gas constituting the so-produced stellar wind bow shock is 
described by the following non-ideal MHD equations, 
\begin{equation}
	   \frac{\partial \rho}{\partial t}  + 
	   \bmath{\nabla}  \cdot \big(\rho\bmath{v}\Big) =   0,
\label{eq:mhdeq_1}
\end{equation}
\begin{equation}
	   \frac{\partial \bmath{m} }{\partial t}  + 
           \bmath{\nabla} \cdot \Big( \bmath{m} \textcolor{black}{\otimes} \bmath{v}  + \bmath{B} \textcolor{black}{\otimes} \bmath{B} + \bmath{\hat I}p \Big) 
            =   \bmath{0},
\label{eq:mhdeq_2}
\end{equation}
\begin{equation}
	  \frac{\partial E }{\partial t}   + 
	  \bmath{\nabla} \cdot \Big( (E+p)\bmath{v}-\bmath{B}(\bmath{v}\cdot\bmath{B}) \Big)  
	  = \Phi(T,\rho),
\label{eq:mhdeq_3}
\end{equation}
and,
\begin{equation}
	  \frac{\partial \bmath{B} }{\partial t}   + 
	  \bmath{\nabla} \cdot \Big( \bmath{v}  \textcolor{black}{\otimes} \bmath{B} - \bmath{B} \textcolor{black}{\otimes} \bmath{v} \Big)  =
	  \bmath{0},
\label{eq:mhdeq_4}
\end{equation}
where $\bmath{B}$ is the magnetic field vector. Additionally, $\rho$ is the mass 
density, $\bmath{m}=\rho\bmath{v}$ stands for the linear momentum vector, 
$\bmath{\hat I}$ is the identity matrix, the quantity $p$ stands for the thermal 
pressure, $v$ is the gas velocity and,  
\begin{equation}
	E = \frac{p}{(\gamma - 1)} + \frac{ \bmath{m} \cdot \bmath{m} }{2\rho} 
	    + \frac{ \bmath{B} \cdot \bmath{B} }{2},
\label{eq:energy}
\end{equation}
if the total energy of the system, respectively. In the above relations, 
$\gamma=5/3$ is the adiabatic index. 
Optically-thin radiative cooling and heating of the gas are represented by 
the source term $\Phi(T,\rho)=n_{\rm H}\Gamma(T)-n_{\rm H}^{2}\Lambda(T)$, 
where $\Lambda(T)$ and $\Gamma(T)$ stand for the gas energy losses and 
photoheating, and where $T = \mu  m_{\mathrm{H}}  p / k_{\rm{B}} \rho$ 
is the gas temperature. These cooling laws of photoionized gas is described 
in great detail in~\citet{meyer_2014bb}.  
%

The MHD equations are solved with the {\sc pluto} code~\citep{mignone_apj_170_2007,migmone_apjs_198_2012,vaidya_apj_865_2018} 
using a $2^{\rm nd}$-order dimensionally unsplit scheme with linear reconstruction 
and the HLL Riemann solver~\citep{hll_ref}. 
The divergence-free condition is controlled using the eight-wave formulation 
by~\citet{Powell1997}. 
The simulations timesteps are controlled by the Courant-Friedrich-Levy condition, that is initially set 
to $C_{\rm cfl}=0.1$.

\subsection{Initial conditions}
\label{sect:method_ic}

%
%
We adopt a Cartesian coordinate system and a $[-0.8;0.8]
\times[-0.8;0.8]\times[-0.2;0.4]\, \mathrm{pc}^{3}$ 
computational domain that is mapped with a uniform grid made of 
$512\times512\times192$ grid 
zones\footnote{
\textcolor{black}{
Our spatial resolution of $3.125\times 10^{-3}\, \mathrm{pc}\, \mathrm{zone}^{-1}$ is 
therefore coarser by an \textcolor{black}{order} of magnitude than that 
of~\citet{meyer_2014a} and~\citet{vanmarle_aa_561_2014}, which are 
$2.25\times 10^{-4}\, \mathrm{pc}\, \mathrm{zone}^{-1}$ and $7.8125\times 10^{-4}\, \mathrm{pc} 
\, \mathrm{zone}^{-1}$, respectively. 
}
}. 
The stellar wind is launched in spherical wind zone of radius $r_{\rm in}=20$ cells 
centered onto the origin of the domain ($0,0,0$), where the wind properties are 
imposed. Particularly, the stellar wind density reads, 
\begin{equation}
	\rho_{w}(r) = \frac{ \dot{M} }{ 4\pi r^{2} v_{\rm w} },
    \label{eq:wind}
\end{equation}
with $\dot{M}$ the mass-loss rate of the star, $r$ the radial distance to the 
origin and $v_{\rm w}$ the terminal wind velocity. 
The simulations are conducted in the reference frame of the star, and our 
method is therefore the 3D Cartesian equivalent of the standard 
cylindrically-symmetric manner of modelling the wind-ISM interaction of massive 
stars~\citep{comeron_aa_338_1998,vanmarle_apj_734_2011,vanmarle_aa_561_2014,
green_aa_625_2019}.

We adopt as initial conditions the stellar properties of the runaway red 
supergiant IRC-14414, which mass-loss rate $\dot{M}\approx 10^{-6}\, 
\rm M_{\odot}\, \rm yr^{-1}$, wind velocity $v_{\rm w}\approx 21\, 
\rm km\, \rm s^{-1}$, ISM gas of number density $n_{\rm ISM}\approx3.3\, \rm cm^{-3}$ 
and stellar bulk motion $v_{\star}=50\, \rm km\, \rm s^{-1}$ have been 
measured and constrained in~\citet{Gvaramadze_2013} and in~\citet{meyer_2014a}, 
respectively. 
\textcolor{black}{
Our stellar wind boundary conditions constrained by observations are motivated 
(i) the availability of real data to be compare our simulations to, and (ii) 
the uncertainty existing regarding to the theoretical estimate of mass-loss 
rates of red supergiant stars~\citep{farrell_mnras_494_2020}. 
}
The star is considered as moving into a fully ionized ambient medium, 
produced either by the \hii region of its   previous    OB main-sequence phase, 
or by a neighbouring stellar cluster~\citep{Gvaramadze_2013,meyer_2014a}. 
Consequently, although red supergiant stars are cool objects, the ISM 
temperature is taken to $T_{\rm ISM} \approx 8000\, \rm K$ and gas obeys 
the heating and cooling rules for photoionized plasma detailed 
in~\citet{meyer_2014a}. 
Inflow and outflow boundary conditions are imposed at the faces of 
our Cartesian computational domain.

For the sake of completeness, a stellar magnetic field $B_{\star}$ is imposed 
at the inner wind boundary, in addition to the wind, as a Parker 
spiral~\citep{parker_paj_128_1958,weber_apj_148_1967}. It is made of a radial component, 
\begin{equation}
	B_{\rm r}(r) = B_{\star} \Big( \frac{R_{\star}}{r} \Big)^{2},
    \label{eq:Br}
\end{equation}
and of a toroidal component, 
\begin{equation}
	B_{\phi}(r) = \textcolor{black}{ B_{\rm r}(r) }
	\Big( \frac{ \textcolor{black}{v_{\phi}(\theta)} }{ v_{\rm w} } \Big) 
	\Big( \frac{ r }{ R_{\star} }-1 \Big),
    \label{eq:Bphi}
\end{equation}
\textcolor{black}{
with, 
\begin{equation}
	v_{\phi}(\theta) = v_{\rm rot} \sin( \theta ),
\label{eq:Vphi}
\end{equation}
}
respectively, where $R_{\star}=1000\, \rm R_{\odot}$ is the stellar 
radius~\citep{dolan_apj_819_2016}, $v_{\rm rot}=5\, \rm km\, \rm s^{-1}$ 
is the angular velocity at the stellar equator of the runaway 
supergiant star Betelgeuse~\citep{kervella_aa_609_2018} and 
$B_{\star}=0.2\, \rm G$ is its stellar surface magnetic 
field~\citep{vlemmings_aa_434_2005}, respectively. The poloidal component of the 
stellar magnetic field is initially set to $B_{\theta}=0\, \rm G$. 
\textcolor{black}{
The rotation axis of the star, is, for the sake of simplicity, taken to be 
the $Oz$ axis. Hence, our problem possesses three characteristic direction: 
the rotation axis ($Oz$), the direction of stellar motion arbitrarily shifted 
along the $Ox$ and $Oy$ directions by $\leq 10\degree$ each (see 
Section~\ref{sect:results_grid}) and the direction of the large-scale 
ISM magnetic field, shifted from that of stellar motion by 
$\theta_{\rm mag}$ (Table~\ref{tab:models}). 
}
Considering the   lack of observational data   regarding to the magnetic 
field of IRC-10414, the rotation and magnetic properties in our models 
have been taken to that of the well-studied evolved cool stars 
Betelgeuse~\citep{kervella_aa_609_2018} and other late-type giant stars 
such as Mira~\citep{vlemmings_aa_394_2002,vlemmings_aa_434_2005}, respectively. 
Following studies tailored to the M-typed star V374 Peg, Proxima Centauri and 
LHS 1140, we scale the toroidal component of the stellar magnetic field to that of the Sun~\citep{herbst_apj_897_2020,baalmann_aa_650_2021}. 
We refer the reader further interested on our magnetised stellar 
wind boundaries to the works of~\citet{chevalier_apj_421_1994,
rozyczka_apj_469_1996,garciasegura_apj_860_2018,garciasegura_apj_893_2020}.

The organised, large-scale magnetization of the ISM is modelled 
by $\vec{B}_{\rm ISM}$ and chosen to lay in a plane described 
by its inclination with respect to the direction of stellar motion. 
We perform several simulations with changing inclination angle $\theta_{\rm mag}$ 
of the local vector magnetic field $\vec{B}_{\rm ISM}$ with respect to the direction 
of stellar motion $-\vec{v}_\star$. \textcolor{black}{In most simulations that we run, 
the inflow is not parallel to the $Oz$ axis (see Section~\ref{sect:results_grid}).}
The strength of the ISM magnetic field is taken to the standard value of 
$B_\mathrm{ISM}=7\, \mu \rm G$ for the warm phase of the 
ISM~\citep{vanmarle_aa_561_2014,vanmarle_584_aa_2015,meyer_mnras_464_2017}. 
Finally, the advection equation, 
\begin{equation}
	\frac{\partial (\rho Q ) }{\partial t } +  \bmath{ \nabla } \cdot  ( \bmath{v} \rho Q ) = 0,
\label{eq:tracer}
\end{equation}
allows us to differentiate stellar wind from ISM materials and to trace the mixing 
of those stellar wind material into the forming gas nebula. The tracer $Q$ is initially 
set to $Q(r\le\,r_{\rm in})=1$ in the wind and to $Q(r>r_{\rm in})=0$ in the ISM, 
respectively. 
Table~\ref{tab:models} summarises the simulation models in our study.


\section{Results}
\label{sect:results}

In this section, we describe and analyse the structure of MHD bow shocks forming 
around a runaway red supergiant star. 
We pay a particular attention to the study of the bow shock stability as a function of the 
angle $\theta_\mathrm{mag}$ between the direction of motion of the star and that of the local 
magnetic field.

\begin{figure}
        \centering
        \includegraphics[width=0.4\textwidth]{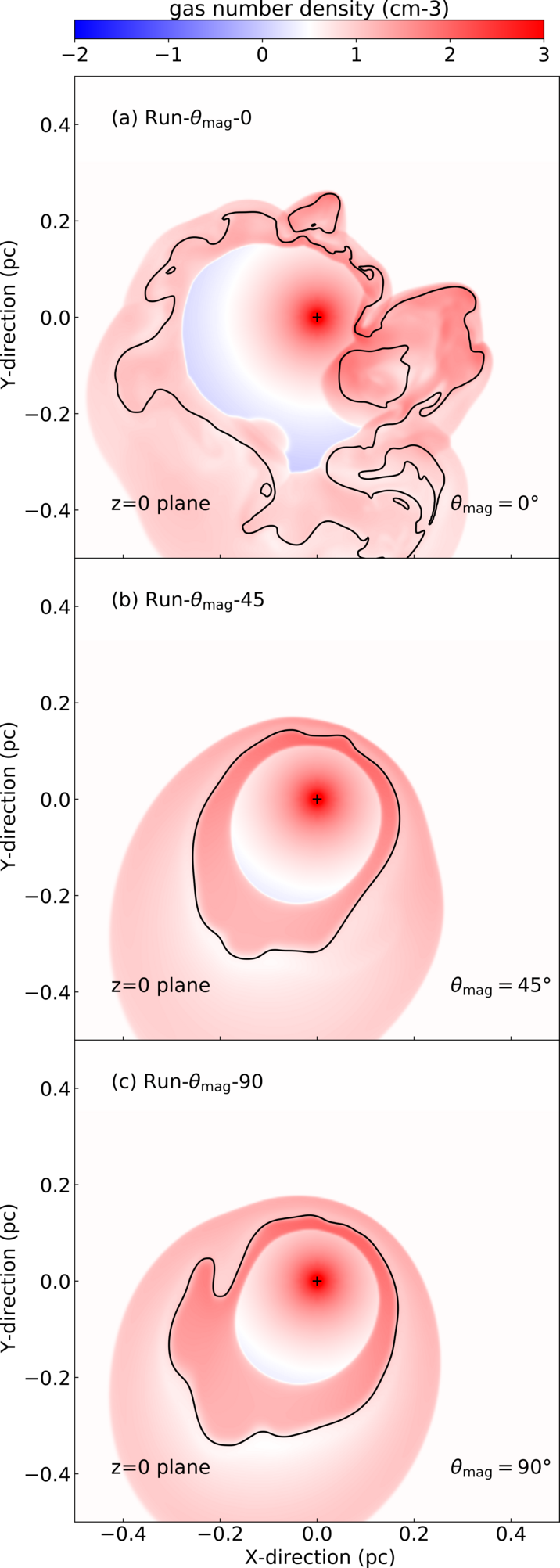}
        \caption{
        Density fields in the $z=0$ plane of models Run-$\theta_\mathrm{mag}$-0, 
        Run-$\theta_\mathrm{mag}$-45 and Run-$\theta_\mathrm{mag}$-90. The simulations 
        assume different angles between the direction of motion of the star and that of the local 
        magnetic field, with $\theta_{\rm mag}=0\degree$ (a), $\theta_{\rm mag}=45\degree$ 
        (b) and $\theta_{\rm mag}=90\degree$ (c), respectively. 
        The black contour is the location in the bow shock where the gas is made of 
        equal proportion of stellar wind and ISM material ($Q=0.5$). 
        The black cross marks the position of the star. 
        }
        \label{fig:density_horizontal1}  
\end{figure}

\begin{figure*}
        \centering
        \includegraphics[width=0.85\textwidth]{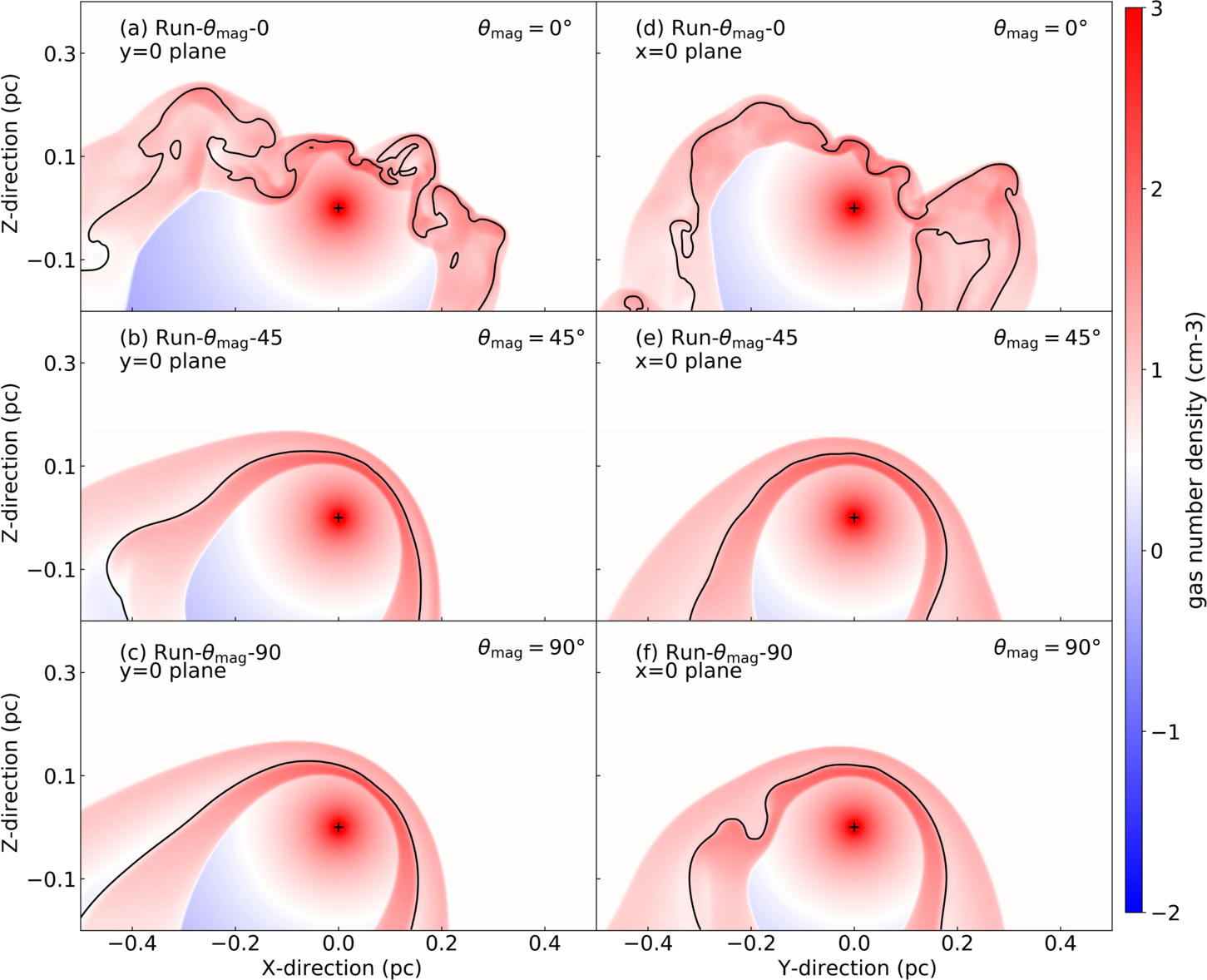}
        \caption{
        Density fields in the $x=0$ (left) and $y=0$ (right) planes of models 
        Run-$\theta_\mathrm{mag}$-0, Run-$\theta_\mathrm{mag}$-45 and 
        Run-$\theta_\mathrm{mag}$-90, assuming different angle between the 
        direction of motion of the star and that of the local magnetic field,  
        $\theta_{\rm mag}=0\degree$ (top), $\theta_{\rm mag}=45\degree$ 
        (middle) and $\theta_{\rm mag}=90\degree$ (bottom), respectively. 
        The black contour is the location of \textcolor{black}{the tangential 
        discontinuity} \textcolor{black}{(where we have equal proportion 
        of stellar wind and ISM material, i.e., $Q=0.5$)}. 
        The black cross marks the position of the star. 
        }
        \label{fig:density_vertical}  
\end{figure*}

\subsection{Bow shock stability as a result of the ISM magnetic field direction}
\label{sect:results_field}

In Fig.~\ref{fig:density_horizontal1} we display the number density field 
(in $\rm cm^{-3}$) in the $z=0$ plane   for   models Run-$\theta_{\rm mag}$-0, 
Run-$\theta_{\rm mag}$-45 and Run-$\theta_{\rm mag}$-90, assuming different 
angles between the direction of motion of the star and that of the local 
magnetic field $\theta_{\rm mag}=0\degree$ (a), $\theta_{\rm mag}=45\degree$ 
(b) and $\theta_{\rm mag}=90\degree$ (c), respectively.
In each panel, the black contour marks the region in the astrosphere  
  at which the mass contribution of stellar and ISM balance   , 
respectively~\cite{meyer_mnras_464_2017}. 
The black cross marks the position of the star. 
The model Run-$\theta_{\rm mag}$-0 with ISM magnetic field direction parallel 
to that of the stellar motion recovers the hydrodynamical limit and produces an  
unstable bow shock exhibiting a very distorted, ragged and clumpy layer of 
shocked ISM material, consistent with the previous 3D hydrodynamical simulations 
of stellar wind bow shocks of~\citet{blondin_na_57_1998}. The \textcolor{black}{astropause}, 
\textcolor{black}{i.e. a tangential discontinuity\footnote{Also called 
astropause, tangential 
discontinuity, wind-ISM interface, or pressure equilibrium surface}} 
(black line), is no more path-connected as a result of the turbulent velocity 
and magnetic fields in it, producing regions of ISM gas engulfed into the layer 
of stellar wind (Fig.~\ref{fig:density_horizontal1}a).

\begin{figure*}
        \centering
        \includegraphics[width=0.9\textwidth]{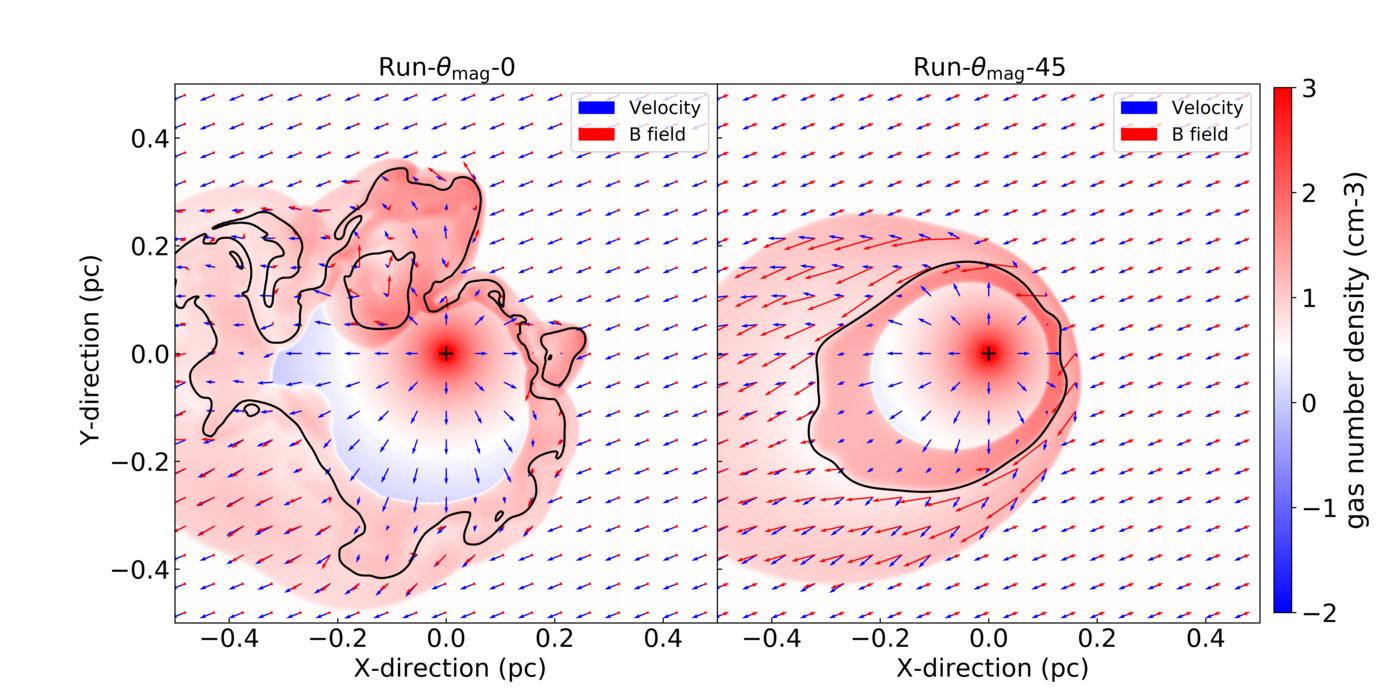}
        \caption{
        Distribution of the gas density in the $z=0$ plane of our bow shock 
        models Run-$\theta_\mathrm{mag}$-0 (left) and Run-$\theta_\mathrm{mag}$-45 
        (right). 
        The velocity and magnetic field vector fields are represented with blue 
        and red arrows, respectively. 
        \textcolor{black}{ 
        Differences between the vector directions are in the plane normal to the 
        figures. 
        }
        The black contour is the location of \textcolor{black}{the tangential 
        discontinuity} ($Q=0.5$). 
        Gas number density is plotted in the logarithmic scale.  
        }
        \label{fig:vectors}  
\end{figure*}

The stand-off distance of the bow shock~\citep{baranov_sphd_15_1971} is 
defined as, 
\textcolor{black}{
\begin{equation}
	R(0) = \sqrt{ \frac{ \dot{M} v_{\mathrm{w}} }
	             { 4 \pi \rho_{\mathrm{ISM}} v_{\star}^{2} } }, 
    \label{eq:Ro}
\end{equation}
where $\rho_{\mathrm{ISM}}=\mu n_{\mathrm{ISM}}  m_{\rm H}$, $n_{\mathrm{ISM}}$ standing 
for the gas density of the ambient medium, $\mu$ the ISM mean molecular weight 
and $m_{\rm H}$ the proton mass. 
}
It is measured   from the simulation data   as    $R(0)\approx 0.14\, \rm pc$ 
(Fig.~\ref{fig:density_horizontal1}a-c). 
\textcolor{black}{
The stand-off distance lies on the inflow line, as~\citet{baranov_sphd_15_1971} 
determined it using an incompressible irrotational fluid model, as 
does implicitly~\citet{wilkin_459_apj_1996} who also assumes a thin-shell 
approach to the overall bow shock structure. 
}
The models with $\theta_{\rm mag}=45\degree$ (Fig.~\ref{fig:density_horizontal1}b) 
and $\theta_{\rm mag}=90\degree$ (Fig.~\ref{fig:density_horizontal1}c) have 
smoother and stable bow shock structures. Because of the extra stress and 
pressure provided by the ISM magnetic field, the layers of \textcolor{black}{astropause} 
and the forward shock are stabilised, see also~\citet{vanmarle_aa_561_2014},  
although the nebula still reveals traces of large scale eddies, repetitively  
\textcolor{black}{advected from} the apex to the tail of the bow shock (black line in 
Fig.~\ref{fig:density_horizontal1}c). 
The layer of shocked ISM material is enlarged and the post-shock density 
at the forward shock is dimmer~\citet{meyer_mnras_464_2017}.

In Fig.~\ref{fig:density_vertical} we plot the number density field 
(in $\rm cm^{-3}$) in the $y=0$ (left) and $x=0$ (right) planes in our 
models Run-$\theta_{\rm mag}$-0 (top) and Run-$\theta_{\rm mag}$-45 (middle) 
and Run-$\theta_{\rm mag}$-90 (bottom), assuming different 
angle between the direction of motion of the star and that of the local 
magnetic field $\theta_{\rm mag}=0\degree$ (a,d), $\theta_{\rm mag}=45\degree$ 
(b,e) and $\theta_{\rm mag}=90\degree$ (c,f), respectively.
As described above, the model   with $\theta_{\rm mag} = 0\degree$   
is clearly unstable and turbulent, with several knots and clumps overpassing 
the stand-off distance. This has been revealed in the 3D hydrodynamical 
simulations of the Rayleigh-Taylor and Vishiniac-unstable 
stellar wind bow shock modelled of~\citet{blondin_na_57_1998}. 
The figures corresponding to the model   with $\theta_{\rm mag} = 90\degree$   
highlight that the \textcolor{black}{astropause} can have different stability 
properties can be different along distinct cross-sections such as the 
$Oxz$ and $Oyz$ planes. 
\textcolor{black}{This induces} projection effects of the nebula's emission onto 
the plane of the sky, according to the viewing angle it is observed from 
(our Section~\ref{sect:emissionmaps}).

Fig.~\ref{fig:vectors} further illustrates the density fields in 
the plane $Oxy$ of models Run-$\theta_{\rm mag}$-0 (left) and 
Run-$\theta_{\rm mag}$-90 (right). The overplotted vectors fields 
highlight the gas velocity (blue) and the magnetisation (red) of 
the plasma. 
On  e  sees the isotropic stellar wind flowing from the central 
black cross marking the position of the star, to the 
\textcolor{black}{tangential discontinuity} (thin black line). 
The size of the vectors scale with the magnitude of the field 
they   represent.   
Hence, the Parker magnetic field in the stellar wind, 
several orders of magnitude smaller than that of the ambient 
ISM, are not visible at the naked eye in these figures. 
The velocity field is reduced in the post-shock region at the 
forward shock, as testifies the smaller blue arrows, which  
re-accelerate in the winged region of shocked ISM behind the 
apex of the astrosphere, see also our discussion on the Mach number 
in Section~\ref{sect:Mach}. 
Similarly, the ISM magnetic field is compressed in the wings of the 
bow shock, filled with shocked ISM material (right panels of 
Fig.~\ref{fig:vectors}).

\begin{figure}
        \centering
        \includegraphics[width=0.45\textwidth]{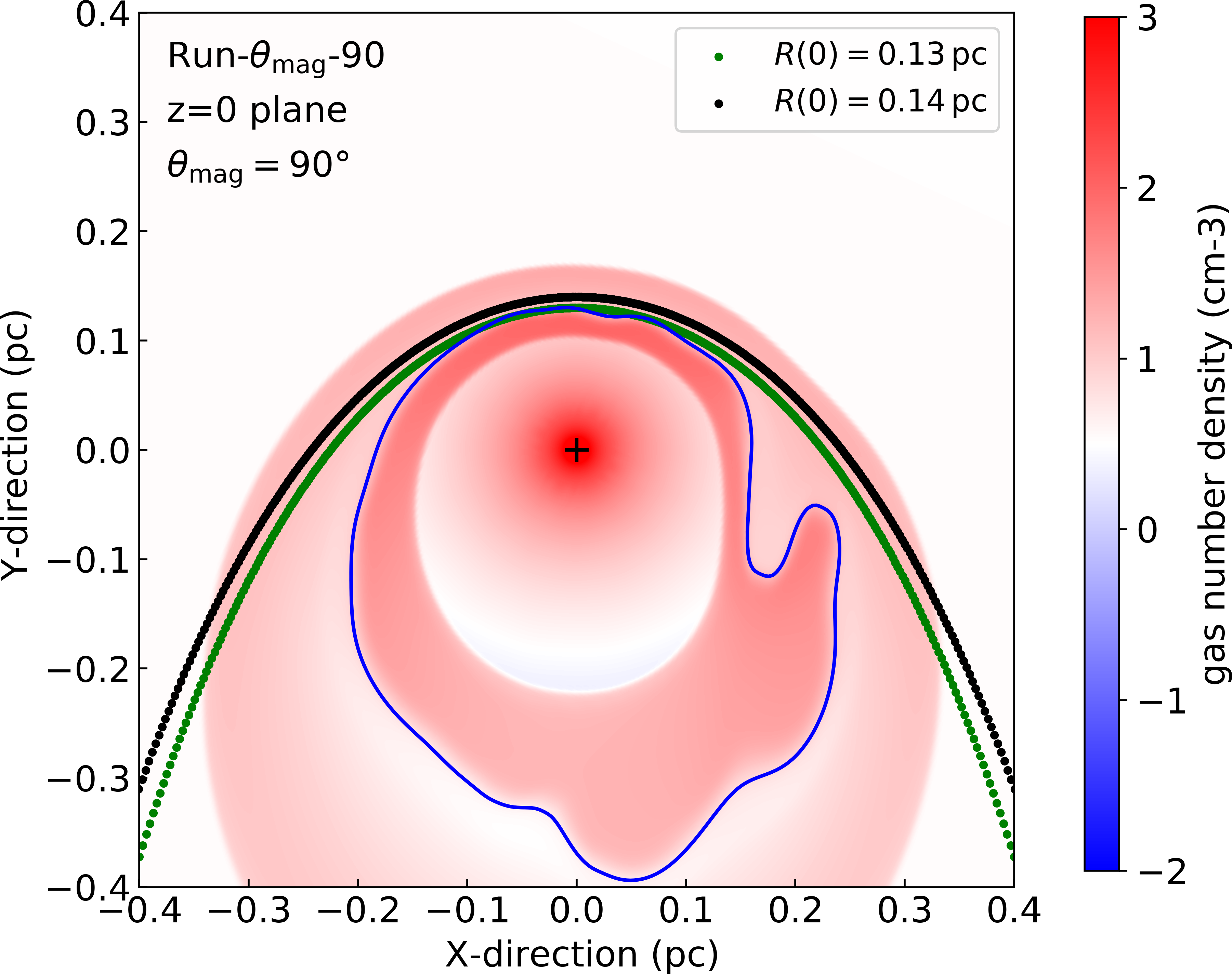}
        \caption{
        Comparison between Wilkin's analytic solution~\citep{wilkin_459_apj_1996} 
        for the overall shape of the bow shock model Run-$\theta_{\rm mag}$-90. 
        Wilkin's solution is given for two stand-off distances, the observed 
        one (black) and that fitting the \textcolor{black}{astropause} of the bow shock 
        nebula (green). 
        }
        \label{fig:wilkin_solution}
\end{figure}

Fig.~\ref{fig:wilkin_solution} compares the analytic solution 
of~\citet{wilkin_459_apj_1996} with the density field 
(in $\rm cm^{-3}$) of the bow shock in our simulation model 
Run-$\theta_{\rm mag}$-90. The figure displays a cross-section 
through the plane $Oxy$ and Wilkin's solution is added as colored
coutours representing the approximation for the morphology of 
an infinitely thin bow shock. It reads, 

\begin{equation}
	R({\theta}) = R(0) \csc(\theta) \sqrt{ 3(1-\theta )\cot(\theta) },
\label{eq:wilkin}
\end{equation}

with $\theta$ is the angle between the direction of stellar motion  in degrees and $R(0)$ the stand-off distance (Eq.~\ref{eq:Ro}).
The black contour   corresponds to   the stand-off distance derived from the 
observations of surroundings of IRC-10414~\citep{Gvaramadze_2013}, 
while the green contour fits the \textcolor{black}{astropause} (marked with 
blue thin contour). 
A couple of remarks naturally arise from this comparison. First, 
the analytic approximation for bow shocks around red supergiant 
stars matches well the overall circumstellar structure. It is 
therefore suitable, e.g. for observers constraining the geometry 
of observed nebulae. However, it does not corresponds to a characteristic 
layer of the bow shock, such as the \textcolor{black}{tangential discontinuity}, 
for angles $\theta\ge20$-$30\degree$. Its applicability is therefore limited 
to the apex of the bow shock~\citep{blondin_na_57_1998,mohamed_aa_541_2012}.

\begin{figure*}
        \centering
        \includegraphics[width=0.75\textwidth]{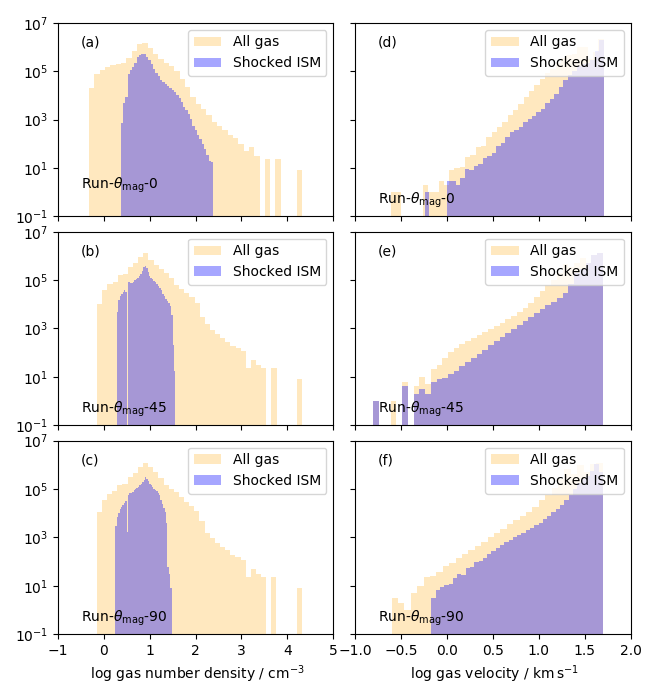}
        \caption{
        Distribution of the gas density (left panels) and gas velocity (right panels) 
        in our bow shock models Run-$\theta_\mathrm{mag}$-0 (top, a,d), 
        Run-$\theta_\mathrm{mag}$-45 (middle, b,e) and 
        Run-$\theta_\mathrm{mag}$-90 (bottom, c,f). 
        The histograms distinguish between the all gas in the bow shock nebulae (yellow) 
        and the shocked ISM material (blue). 
        Gas number density and velocity are plotted in the logarithmic scale.  
        }
        \label{fig:hist}  
\end{figure*}

\subsection{Bow shock internal properties}
\label{sect:Mach}

Fig.~\ref{fig:hist} shows the distribution of the gas density (right) and gas velocity (left), for both all gas in the bow shock nebulae (yellow) and for the shocked ISM gas component (blue), respectively. 
Figures are displayed   for different angles, namely,   $\theta_{\rm mag}=0\degree$ 
(top), $\theta_{\rm mag}=45\degree$ (middle) and $\theta_{\rm mag}=90\degree$ 
(bottom). Gas number density (in $\rm cm^{-3}$) and velocity (in $\rm km\, \rm s^{-1}$) 
are plotted in   log-  scale, while the distribution represents the 
number of grid zones which is a measure of the volume. 
  Inspection of the left panels reveals   that the inclination of the ISM magnetic field $\theta_{\rm mag}$ induces notable differences in the bow shock density distribution between the models, see e.g. models with 
$\theta_{\rm mag}=0\degree$ (Fig.~\ref{fig:hist}a) and with 
$\theta_{\rm mag}=45\degree$ (Fig.~\ref{fig:hist}b). 
Indeed, the unstable model in the hydrodynamical limit (Fig.~\ref{fig:hist}a) has a 
density distribution which is globally   larger   than in the other models (Fig.~\ref{fig:hist}b,c). 
This illustrates the well-know effect of magnetic fields in dimming the gas 
density in post-shock region at the forward shock~\citep{meyer_mnras_464_2017}. 
Note that density distribution peaks   are   similar in all three models 
(Fig.~\ref{fig:hist}a-c) as each bow shock model is driven by the same star. 
Not much difference is noticeable between the models with 
$\theta_{\rm mag}=45\degree$ and $\theta_{\rm mag}=90\degree$, 
suggesting also that their respective maximal emissivity $\propto n^{2}$ 
might not greatly differ compare to the case with 
$\theta_{\rm mag}=0\degree$ (see Section~\ref{sect:emissionmaps}).

The right panels display histogram for the gas velocity distributions in the 
astrospheres. The similar upper part of all distributions (Fig.~\ref{fig:hist}d-f) 
in each panels comes from the fact that all simulations have the same central 
star expelling the same winds, moving at the same space velocity 
$v_{\star}=50\, \rm km\, \rm s^{-1}$ through the same ISM. Only mild differences   are found   in the low-velocity region of the two distributions, because of   the   changes in the velocity field in the post-shock region at the 
\textcolor{black}{forward (bow) shock and at the reverse (termination) shock}, 
respectively 
The more the flow is laminar in the shocked ISM (the less the tangential  
discontinuity is unstable) the lower will be the minimum gas velocity 
(Fig.~\ref{fig:hist}e). This may have observational consequences, e.g. 
on atomic and molecular line emissions of the nebula, which is much beyond 
the scope of this study.

\begin{figure*}
        \centering
        \includegraphics[width=0.9\textwidth]{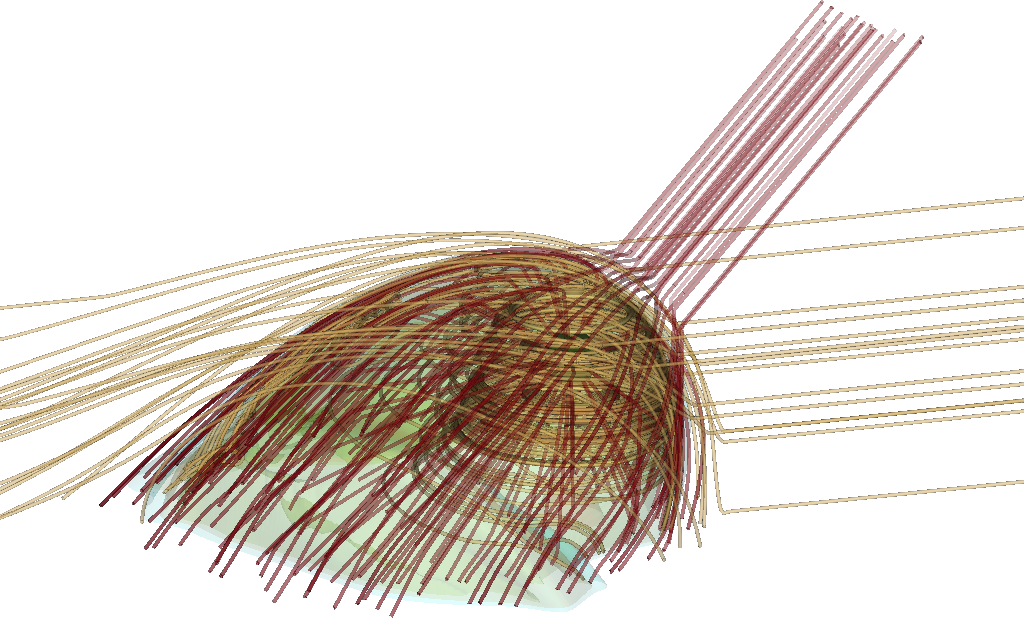}
        \caption{
        Rendering of the structure of the 3D MHD bow shock in our model 
        Run-$\theta_\mathrm{mag}$-45, forming around a runaway red supergiant 
        star moving into an constant ambient medium in which the direction of 
        stellar motion and local magnetic field direction make an angle 
        of $\theta_\mathrm{mag}=45\degree$. 
        The transparent clipped surfaces are number density isocontours. 
        The yellow and red lines are gas velocity and magnetic field 
        streamlines, respectively. 
        %
        %
        }
        \label{fig:3D}  
\end{figure*}

In Fig.~\ref{fig:3D} we further illustrate the internal structure of 
the stable bow shock in our model Run-$\theta_{\rm mag}$-$45$. The plot 
is a three-dimensional rendering of the astrophere around the runaway red 
supergiant star IRC-10414. Transparent clipped number density surfaces 
render the bow shock structure and streamlines mark the magnetic field (yellow) 
and velocity fields (red), respectively. 
It shows how the stellar magnetic field, initially made of a Parker spiral 
transported with the stellar wind in the equatorial plane of the star, fills the 
entire region of freely-expanding, accelerated unshocked stellar wind. 
These field lines do not reconnect with the ISM magnetic field lines that  
penetrate the forward shock and 
stabilise the \textcolor{black}{astropause}, but rather expand as a wider spiral 
in the \textcolor{black}{tail} of the bow shock. 
The gas streamlines also highlight the isotropically-outflowing stellar wind 
from the inner stellar boundary, which enters the region of shocked wind 
and bends towards the tail of the astrosphere. \textcolor{black}{The ISM gas enters the bow 
shock in its turn and flow along the wind-ISM 
discontinuity.}

\begin{figure*}
        \centering
        \includegraphics[width=0.9\textwidth]{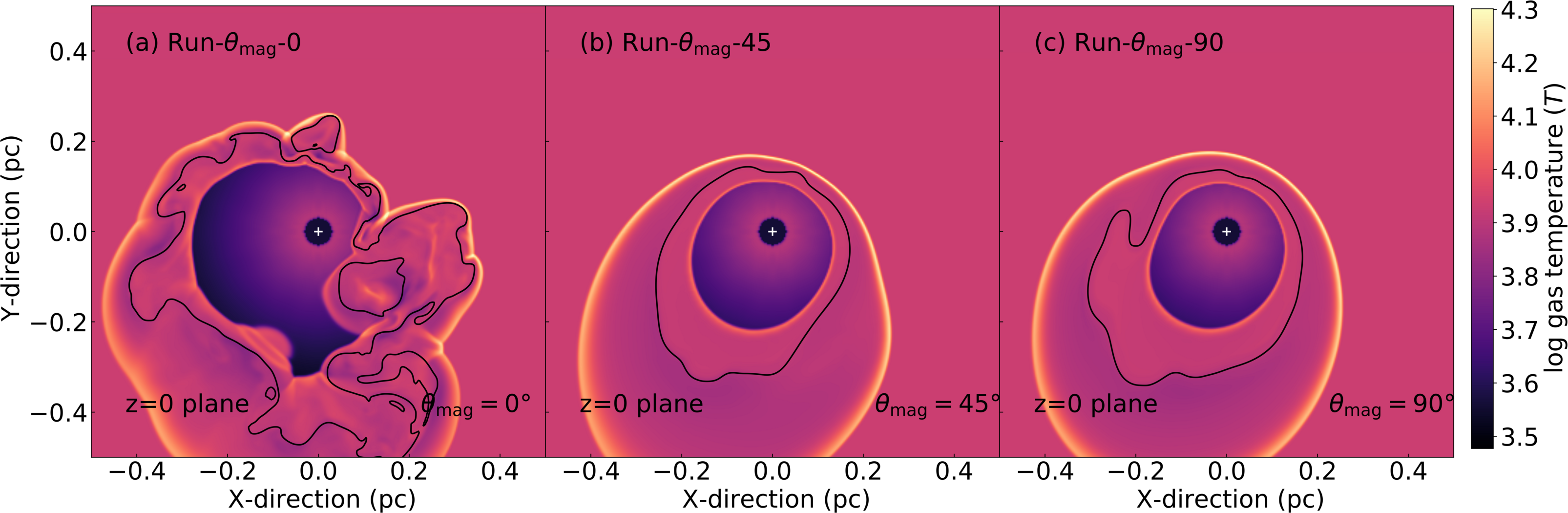} \\ 
        \centering
        \includegraphics[width=0.9\textwidth]{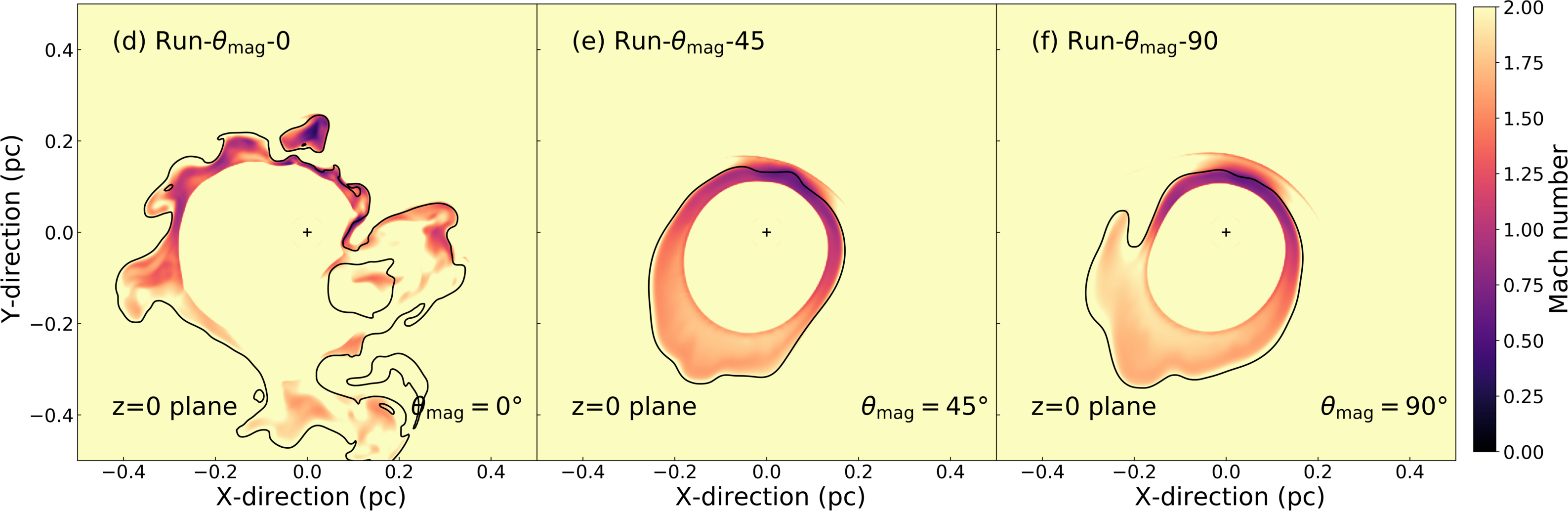}  \\
        \centering
        \includegraphics[width=0.9\textwidth]{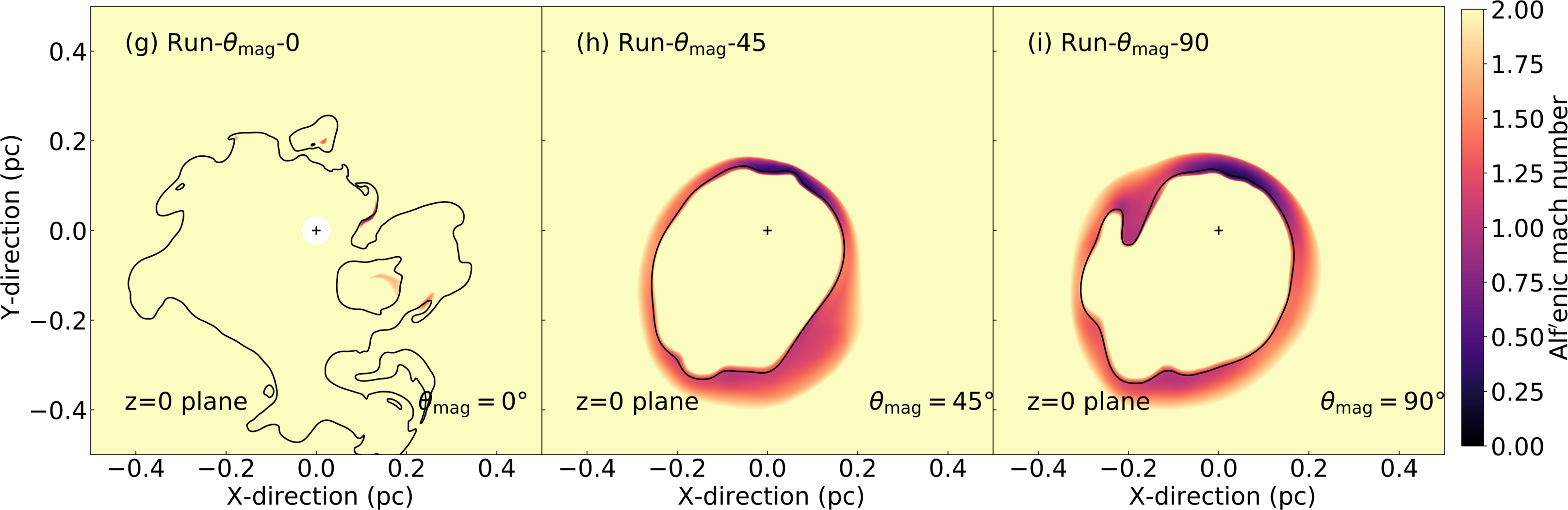} \\
        \centering
        \includegraphics[width=0.9\textwidth]{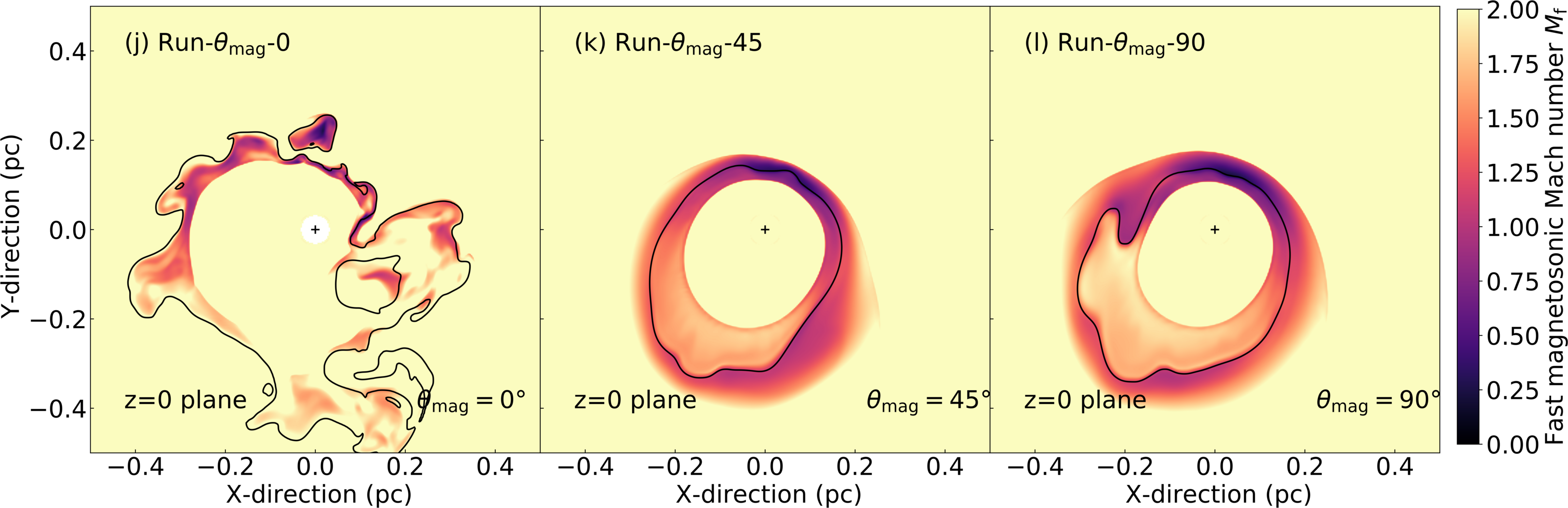} \\
        \caption{
        Temperature $T$ (top panels), sonic Mach number $M$ (middle top panels), 
        Alf\'evnic Mach $M_{\rm A}$ number (middle top panels) and fast magnetosonic 
        Mach number $M_{\rm f}$ (middle bottom panels) in the $z=0$ plane of our 
        bow shock models.  
        Bow shock models are displayed as a function of the angle between the 
        direction of stellar motion and ISM magnetic field, which spans from 
        $\theta_{\rm mag}=0\degree$ (left) to $\theta_{\rm mag}=90\degree$ (right). 
        The black contour is \textcolor{black}{astropause} interface into the 
        bow shock, and on each panel the central black and white crosses marks 
        the position of the star. 
        }
        \label{fig:inclinaison_rho}  
\end{figure*}

Fig.~\ref{fig:inclinaison_rho} displays the temperature field (top panels), 
sonic Mach number field (middle top panels), Alf\'evnic Mach number field (middle 
top panels) and fast magnetosonic Mach number field (middle bottom panels) 
in the $z=0$ plane of our bow shock models, displayed as a function of 
$\theta_{\rm mag}=0\degree$ (left), $\theta_{\rm mag}=45\degree$ (middle) 
and $\theta_{\rm mag}=90\degree$ (right). 
The black contour is \textcolor{black}{astropause} interface of the bow shock, i.e. the 
location of the nebula made of equal proportion of stellar wind and ISM 
material ($Q=0.5$). On each panel the cross marks the position of the star. 
The temperature maps illustrate that the maximum temperature is at 
the bow shock, where it exceeds that at the termination 
shock (Fig.~\ref{fig:inclinaison_rho}a-c), see also~\citet{meyer_mnras_464_2017}.

The series of panels on the Mach number, 
\begin{equation}
	M = \frac{ v }{ c_{\rm s} }, 
    \label{eq:M}
\end{equation}
shows that the gas is supersonic in the free-streaming wind and in the 
unshocked ISM gas, respectively. Indeed, the accelerated stellar wind 
$v_{\rm w}$ is larger than the radially-decreasing temperature by 
adiabatic cooling $c_{\rm s}$, and $v_{\star}=50\, \rm km\, \rm s^{-1}>c_{\rm s}
\approx 10\, \rm km\, \rm s^{-1}$ everywhere in the unperturbed ISM. 
The gas is sub-sonic in the shocked regions, both \textcolor{black}{located downstream 
the termination shock and the bow shock}, as a result of 
the changes in both the post-shock temperature and velocity ($M<1$). 
This permits us to clearly classify the astrosphere of IRC-10414 as a bow shock, 
instead of other circumstellar structures such as bow wave or dust waves 
which can by produced around moving  
stars~\citep{pogorelov_apj_845_2017,henney_2019_arXiv190400343H}. 
Note also that the sonic Mach number increases in the region of 
shocked stellar wind gas in the in wings of the bow shock 
(Fig.~\ref{fig:vectors}e,f).

The Alf\' enic Mach number is defined as,
\textcolor{black}{
\begin{equation}
	M_{\rm A} = \frac{ v }{ v_{\rm A} } 
	= \frac{ v }{ |\vec{B}|/\sqrt{ \rho } }, 
    \label{eq:Ma}
\end{equation}
}
and the fast-magnetosonic Mach number reads as, 
\begin{equation}
	M_{\rm f} = \frac{ v }{ v_{\rm f} }
    \label{eq:Mf}
\end{equation}
with the fast-magnetosonic speed, 
\begin{equation}
	v_{\rm f} = \sqrt{ \frac{1}{2} \Big( ( c_{\rm s}^{2} + v_{\rm A}^{2} ) 
	+ \sqrt{ |( c_{\rm s}^{2} + v_{\rm A}^{2} )^{2} - 4 c_{\rm s}^{2} v_{\rm A}^{2} 
	\cos(\Theta)^{2} | \Big) } } ,
    \label{eq:vf}
\end{equation}
where $\Theta$ is the angle between the gas velocity $\vec{v}$ and magnetic 
field $\vec{B}$ vector fields. 
\textcolor{black}{
Note that the sound speed is the characteristic dynamical quantity of an 
hydrodynamical simulation, whereas the fast-magnetosonic speed is the 
relevant dynamical quantity of an magneto-hydrodynamical model, respectively. 
}
Expanding stellar wind and unperturbed ISM 
are supersonic, super-Alf\' enic and super-fast magnetosonic. 
The values of the sonic Mach number $M$ is lower in the shocked wind region 
than in that of the shocked ISM gas because the post-shock gas velocity at 
the termination shock is smaller than at the forward shock, and because the 
sound speed is larger in that dense region of shocked wind 
(Fig.~\ref{fig:inclinaison_rho}a-c). 
Similarly, the Alf\' enic Mach number $M_{\rm A}$ is smaller in the shocked 
ISM region at the apex of the bow shock than in the layer of shocked stellar wind, 
because of both the changes in compressed magnetic field lines and gas number 
density (Fig.~\ref{fig:inclinaison_rho}h,i). 
%
%
The fast-magnetosonic Mach number $M_{\rm f}$ is below unity in all shocked 
regions of the stable bow shocks but larger than unity in the freely 
streaming stellar wind and in the ambient medium, as described 
in~\citet{scherer_mnras_493_2020}.

\begin{figure}
        \centering
        \includegraphics[width=0.49\textwidth]{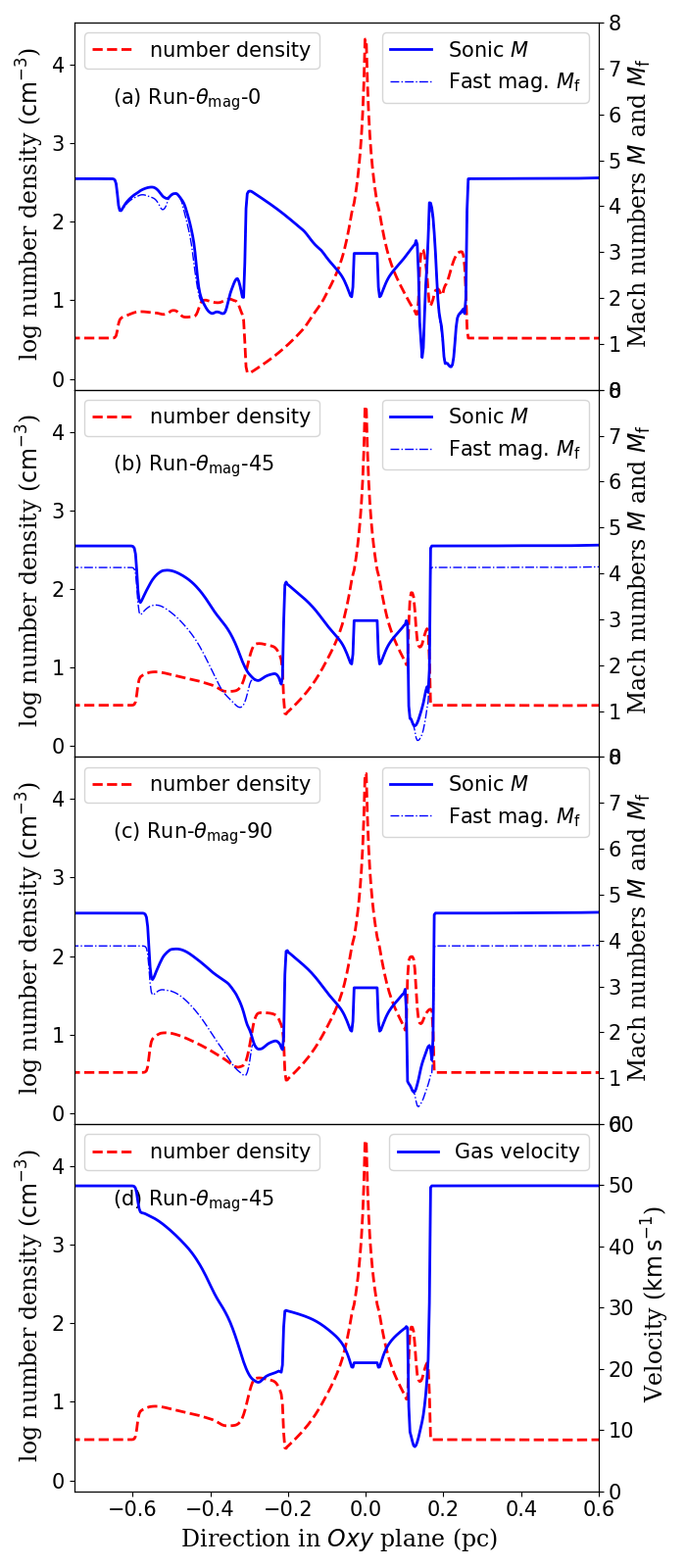}
        \caption{
        Cuts along the $Oy$ axis in our Run-$\theta_\mathrm{mag}$-0 (a), 
        Run-$\theta_\mathrm{mag}$-45 (b) and Run-$\theta_\mathrm{mag}$-90 (c), 
        showing the gas number density, the sonic Mach number $M$ and the fast-magnetosonic 
        Mach number $M_{\rm f}$ profiles. 
        \textcolor{black}{
        The last panel shows the acceleration of the stellar wind in 
        Run-$\theta_\mathrm{mag}$-45 (d). 
        }
        Gas number density is plotted in the logarithmic scale, Mach numbers 
        \textcolor{black}{and gas velocity} in the linear scale.  
        }
        \label{fig:cuts}  
\end{figure}

Fig.~\ref{fig:cuts} shows cross-sections in the number density and Mach 
number fields taken through the maps displayed in Fig.~\ref{fig:inclinaison_rho}, 
measured along the $y=0$ directions. 
The stellar wind density profile is clearly visible at the origin of 
the domain, which decreases $\propto 1/r^{2}$ to values that are lower in the $y<0$ 
part of the figure than in the $y>0$, in the region of the bow shock opposite of 
the direction of stellar motion. 
\textcolor{black}{
The sonic Mach number $M$ profile reflects the decrease of the temperature 
in the stellar wind under the effect of adiabatic cooling 
while the the external photo-heating accelerates the 
stellar wind (Fig.~\ref{fig:cuts}d)}, see also~\citep{meyer_2014a,2014Natur.512..282M}, 
from the inner stellar wind region that is excluded from the computational 
domain, to the termination shock of the bow shock. As expected and above described, 
\textcolor{black}{
the fast magnetosonic Mach number $M_{\rm f}$ number diminishes and is $M_{\rm f}<1$ 
in the post-shock layers of materials located between the termination shock and the bow 
shock (Fig.~\ref{fig:cuts}c). 
}

\begin{figure*}
        \centering
        \includegraphics[width=0.9\textwidth]{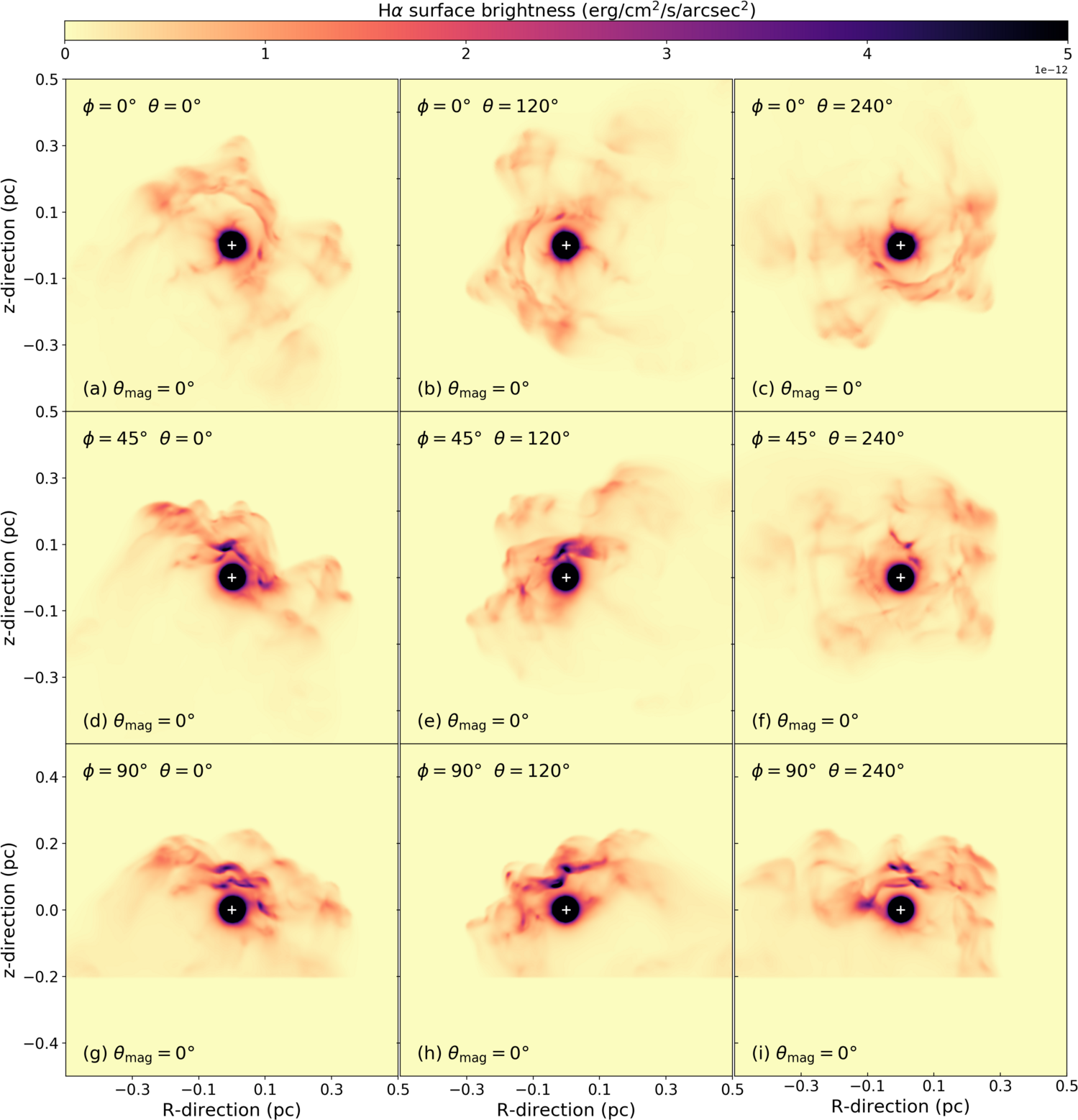} \\
        \caption{
        Emission maps of our bow shock model Run-$\theta_\mathrm{mag}$-0. 
        The figure plots the H$\alpha$ surface brightness 
        (in $\rm erg\, \rm cm^{2}\, \rm s^{-1}\, \rm arcsec^{-2}$). 
        Quantities are calculated excluding the 
        undisturbed ISM and plotted in the linear scale, under several 
        viewing angles $\phi$ and $\theta$, respectively. 
        The white cross marks the position of the star. 
        }
        \label{fig:maps_0}  
\end{figure*}

\begin{figure*}
        \centering
        \includegraphics[width=0.9\textwidth]{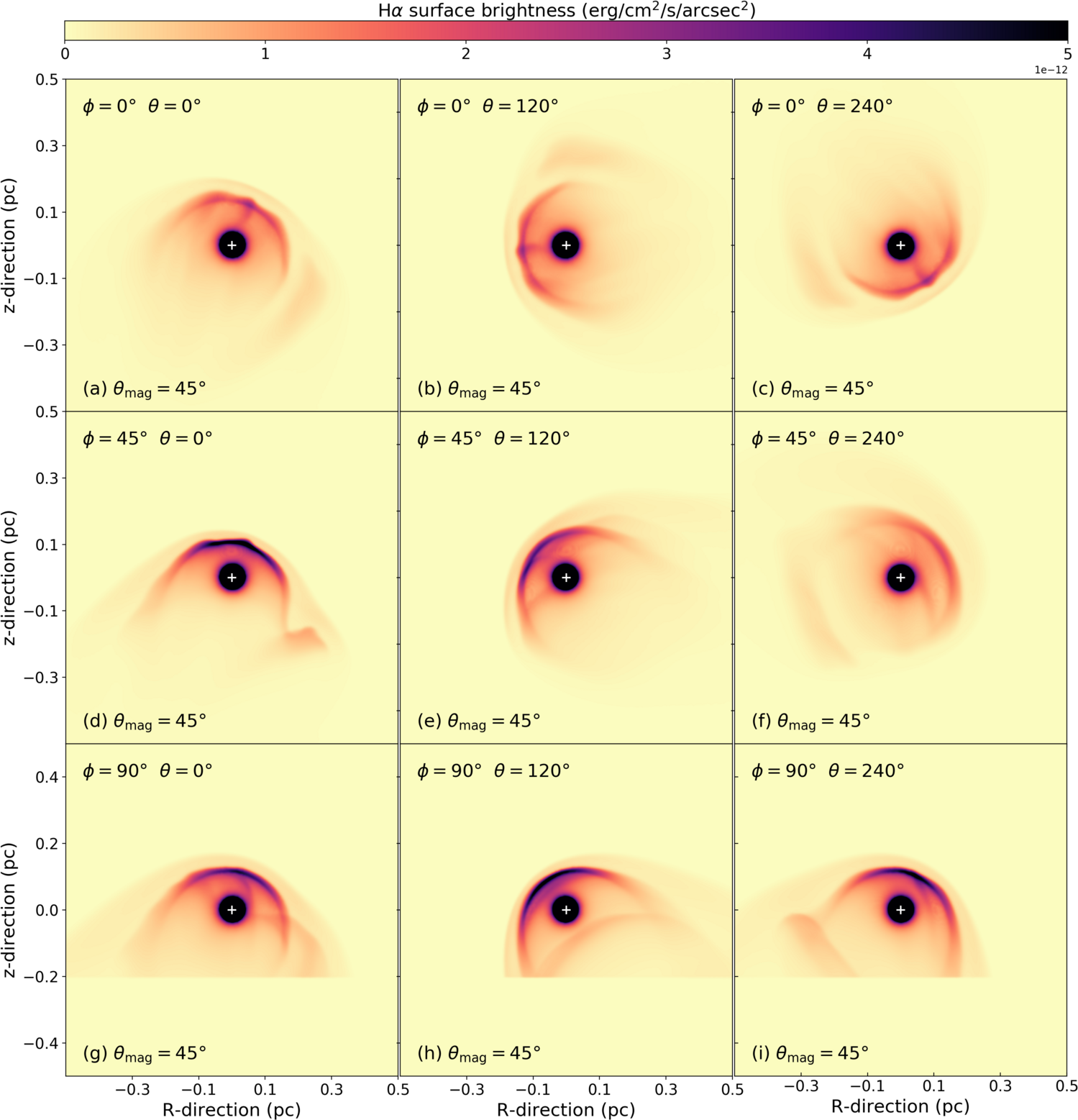} \\        
        \caption{
        Same as Fig.~\ref{fig:maps_0} for our model Run-$\theta_\mathrm{mag}$-45. 
        }
        \label{fig:maps_45}  
\end{figure*}

\begin{figure*}
        \centering
        \includegraphics[width=0.9\textwidth]{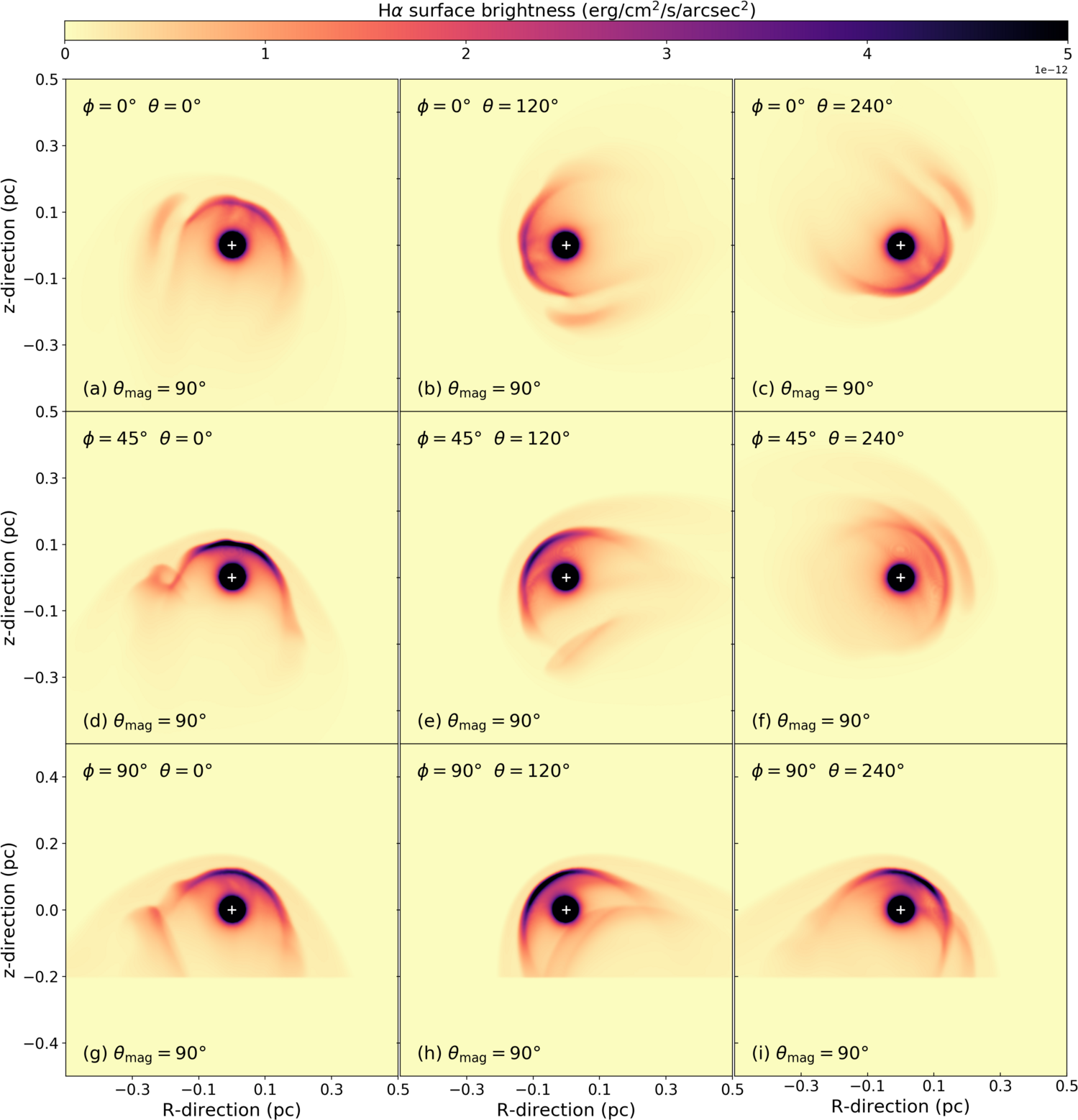} \\        
        \caption{
        Same as Fig.~\ref{fig:maps_0} for our model Run-$\theta_\mathrm{mag}$-90. 
        }
        \label{fig:maps_90}  
\end{figure*}

\subsection{Emission maps}
\label{sect:emissionmaps}

We produce emission maps using the {\sc radmc-3d} code~\citep{dullemond_2012}, 
which permits calculating radiative transfer for analytic prescription of 
emission coefficients such as optical H$\alpha$ optical emission. 
In Fig.~\ref{fig:maps_0}, we plot H$\alpha$ emission maps of our bow 
shock model Run-$\theta_\mathrm{mag}$-0. The figure displays the surface 
brightness $\Sigma$ (in $\rm erg\, \rm cm^{2}\, \rm s^{-1}\, \rm arcsec^{-2}$) 
\textcolor{black}{of a nebula located at a distance of $2\, \rm kpc$}, 
corresponding \textcolor{black}{to the IRC-10414 distance}~\citep{Gvaramadze_2013}. 
The projected emissivity is shown under several considered rotation of the 
polar $\phi$ and azimuthal $\theta$ viewing angles, respectively. 
The white cross marks the position of the runaway star. 
\textcolor{black}{The bow shock of the model Run-$\theta_\mathrm{mag}$-0 
results turbulent. Its exhibits a diffuse nebula morphology, showing several 
filaments and knots dispersed around the central object.} The star is much brighter 
than its surroundings, which raises the question of the saturation 
of observed fluxes and of the screening of red supergiant bow shocks.

A series of two arcs is visible ahead of the star, roughly in the direction 
of stellar motion (Fig.~\ref{fig:maps_0}a). However, the star is not at the 
geometrical centre of the arc and some clumps are much beyond the bow shock 
(Fig.~\ref{fig:maps_0}a-c). Note that this structures are very different 
from the circular ones obtained with 2D hydrodynamical simulations 
in~\citet{meyer_2014bb}. 
The images with $\phi=45\degree$ are brighter than that \textcolor{black}{with $\phi=0\degree$ 
by a factor $2$$-$$3$?}, as expected from previous 2D simulations. \textcolor{black}{This behaviour is because} the line-of-sight 
intercepts more dense material and consequently produces brighter surface 
brightness (Fig.~\ref{fig:maps_0}d-f). \textcolor{black}{The observed star-arc 
distance (called stand-off distance in the observational literature)} varies 
\textcolor{black}{significantly} between models, some of them revealing distorted and clumpy arcs close to 
the star (Fig.~\ref{fig:maps_0}d). The changes in the bow shock morphology 
are important for mild changes in the viewing angle (Fig.~\ref{fig:maps_0}e). 
In these conditions, it is very difficult to distinguish a circumstellar 
structure, and even more uneasy to identify bow shocks, if at all, 
although the nebula has formed and exist (Fig.~\ref{fig:maps_0}). 
%

In Fig.~\ref{fig:maps_45}, we display H$\alpha$ emission maps of our astrosphere 
model Run-$\theta_\mathrm{mag}$-45. The  projected emissivity morphology is very different than in the case with $\theta_{\rm mag}=0\degree$, 
as \textcolor{black}{evident} bow shock morphologies are visible in each panel of the figure. 
The maximum emission originates from the sub-sonic and super-Alf\'enic 
region of shocked stellar wind, much denser than the compressed 
magnetised ISM material, plus a lighter overarching arc tracing the forward 
shock of the bow shock. 
Bright clumps appear in the shocked wind close to the region of maximum 
emission (Fig.~\ref{fig:maps_45}a-c). This \textcolor{black}{fact} reflects the instabilities 
at the interface between wind and ISM material visible in the $Oxy$ plane 
(black line in Fig.~\ref{fig:density_vertical}b). 
These infrared bow shocks are not strictly symmetric with respect to 
the direction of motion of the star, as a consequence of the inclined 
magnetic field direction inducing extended wing-like regions of 
shocked ISM gas of different size (Fig.~\ref{fig:density_vertical}b). 
By projection effects, these regions can produce brighter zones 
wrongly suggesting, from \textcolor{black}{a} purely observational point of view, the
presence of filaments, while it relies large-scale instabilities in 
the \textcolor{black}{astropause} interface. High-resolution simulations would result in 
qualitatively different patterns, and this should be explored in future 
works. This structure can also appear as stripes normal to the direction 
of motion of the runaway star (Fig.~\ref{fig:maps_0}e,f) difficult to 
interpret from the observational point of view. 
As in Fig.~\ref{fig:maps_0} the surface brightness of the astrosphere 
is more significant with $\theta=45\degree$ than with $\theta=0\degree$. 
Finally, note that some models are also very smooth (Fig.~\ref{fig:maps_0}h) 
and fit well with the 2D predictions of~\citet{meyer_2014bb}, see 
Section~\ref{sect:discussion_irc10414}.

In Fig.~\ref{fig:maps_90}, we show H$\alpha$ emission maps for our bow 
shock model Run-$\theta_\mathrm{mag}$-90. The situation is similar to that 
in our model Run-$\theta_\mathrm{mag}$-45. Several shapes arise from 
projection effects, with differences in the maximum emission of the 
circusmtellar material (Fig.~\ref{fig:maps_90}a). Very smooth shells (Fig.~\ref{fig:maps_90}h)   
or irregular (Fig.~\ref{fig:maps_90}d,g,i) patterns of dense shocked 
stellar wind are mostly consequences of the unstable tangential discontinuity. 
Interestingly, the perpendicular orientation of the ISM magnetic field 
can result in series of multiple arcs normal to the direction of 
motion of the star (Fig.~\ref{fig:maps_90}f) as modelled in the 
context of hot OB stars in~\citet{katushkina_MNRAS_465_2017,katushkina_MNRAS_473_2018}, 
\textcolor{black}{ see also~\citet{baalmann_aa_650_2021}. }
\textcolor{black}{Furthermore,} the opening of the bow shock can significantly  differ from a projection model 
to another, with rounder termination shocks (Fig.~\ref{fig:maps_90}f) 
and wider arcs (Fig.~\ref{fig:maps_90}d). 
Overall, our emission maps reveal the importance of projection effects 
in the appearance of stellar wind bow shocks from runaway red supergiant 
stars at H$\alpha$. It supports the use of astrometric measures, such as 
{\sc Gaia} in order to determine the direction of proper motion of runaway 
stars instead of their circumstellar bow shocks, the latter being a source 
of errors resulting from the random orientation of projected astrospheres, 
\textcolor{black}{see~\citet{peri_aa_578_2015}. }



\section{Discussion}
\label{sect:discussion}

In this section, we discuss the limitations of our simulation method. 
As application, we reproduce the bow shocks of two runaway red supergiant stars, 
IRC-10414 and Betelgeuse \textcolor{black}{(see below)}, respectively.
Last, we discuss our findings in the context of nebulae around other massive stars 
and core-collapse supernova remnants.

\subsection{Model limitation}
\label{sect:discussion_caveats}

Our 3D MHD simulations are a leap forward in studying the circumstellar 
medium of runaway late-type (super)giant stars, as it includes both stellar and 
ISM magnetic fields within a three-dimensional framework, completed with radiative 
transfer calculations. 
\textcolor{black}{Therefore, the simulations are} intrinsically more realistic, for example in terms 
of development and damping of instabilities in the bow shock structure, compare 
to that previously conducted in, e.g.~\citet{meyer_2014bb}. Mainly, it is possible 
to explore and study the effects of an ISM magnetic field that is not aligned with 
the direction of stellar motion, which latter configuration was imposed by the 
two-dimensional cylindrical coordinate system used in the other studies of that 
series and in~\citet{vanmarle_aa_561_2014}. 
We study this effect by varying the angle $\theta_{\rm mag}$ between the 
direction of stellar motion and the direction of the ISM magnetic field. Our results 
stress the stabilising role of the ISM magnetic field already noticed in 2D precedent 
works with $\theta_{\rm mag}=0\degree$~\citep{vanmarle_aa_561_2014,meyer_mnras_464_2017}.

Nevertheless, \textcolor{black}{within the stellar wind properties of IRC-10414,} 
the damping of instabilities in stellar wind bow shocks is very efficient in 3D 
only if $\theta_{\rm mag}\ge$ a few degrees, while simulations with 
$\theta_{\rm mag}=0\degree$ recovers the hydrodynamical, unmagnetised limit 
(Fig.~\ref{fig:density_horizontal}). 
These latter results have different stability properties than the 2D models 
of~\citet{meyer_2014bb}. This effect is a noteworthy difference between our 
3D simulations and precedent 2D works 
\textcolor{black}{
in the context of the surroundings of IRC-10414. We interpret this difference 
as originating from the dissimilar coordinate systems used in these two works. 
The first one employs a cylindrical two-dimensional coordinate system which 
possesses a symmetry axis affecting the development of instabilities and 
leading to the complex topology of the bow shock 
apex~\citep{comeron_aa_338_1998,meyer_mnras_464_2017}. The second one considers 
a 3D Cartesian coordinate system, which permits circumventing that 
issue~\citep{mignone_jcoph_270_2014}.
}

The spatial resolution of the simulations has \textcolor{black}{become} an essential parameter  
regarding the development of bow shock instabilities~\citep{vanmarle_aa_561_2014}. 
In that sense, future simulations should consider simulation models with \textcolor{black}{even 
higher} spatial resolution. However, one should keep in mind that fully resolving 
Rayleight-Taylor-based instabilities can not be achieved numerically using Eulerian 
codes~\citep{blondin_na_57_1998,mohamed_aa_541_2012}, and that tending to such 
result would be at \textcolor{black}{prohibitive} numerical costs. 
Furthermore, our models does not include all microphysical mechanisms at work in 
the circumstellar medium of late-type giant stars, such non-ideal MHD processes 
like Ohmic diffusion and Hall effect, known to affect the density of magnetised 
plasmas, chemical reactions at work in the shocked enriched stellar 
winds or acceleration of stellar wind electrons and protons at the termination 
shock of bow shock nebulae~\citep{valle_ApJ_864_2018}. 
Apart from lacking physical processes, our choices \textcolor{black}{regarding} boundary 
and initial conditions for stellar wind and the ISM \textcolor{black}{will be studied in 
a future parameter study}.

Indeed, the present study is tailored to the runaway red supergiant IRC-10414, which surface 
properties have been constrained by means of observations~\citep{Gvaramadze_2013} 
completed with numerical simulations~\citep{meyer_2014a}. Our models, therefore, 
do not include stellar evolution as wind boundary conditions, e.g. time-dependently 
interpolating evolutionary tracks~\citep{brott_aa_530_2011a,2020arXiv200408203S}, 
which would permits to better treat the surroundings of red supergiant 
stars~\citep{mackey_apjlett_751_2012,meyer_2014bb}. \textcolor{black}{On the 
other hand, the stellar evolution} can be useful, to 
model core-collapse supernova remnants from $10$-$20\, \rm M_{\odot}$ massive 
runaway progenitor stars~\citep{katsuda_apj_863_2018}. 
Additionally, the manner of the surrounding medium remains rather simplistic in 
our study as we assume a uniform, warm and ionized ISM. A natural improvement will 
assume a more realistic turbulent, inhomogeneous~\citep{walch_773_apj_2011} and 
clumpy~\citep{baalmann_aa_650_2021} medium. 
This would account for the intrinsic turbulence driven by the stellar winds of 
neighbouring massive stars and/or travelling shock wave from supernova 
remnants~\citep{peters_mnras_467_2017,seifried_apj_855_2018}.

\subsection{Grid-induced effects on bow shock instability}
\label{sect:results_grid}

\begin{figure}
        \centering
        \includegraphics[width=0.4\textwidth]{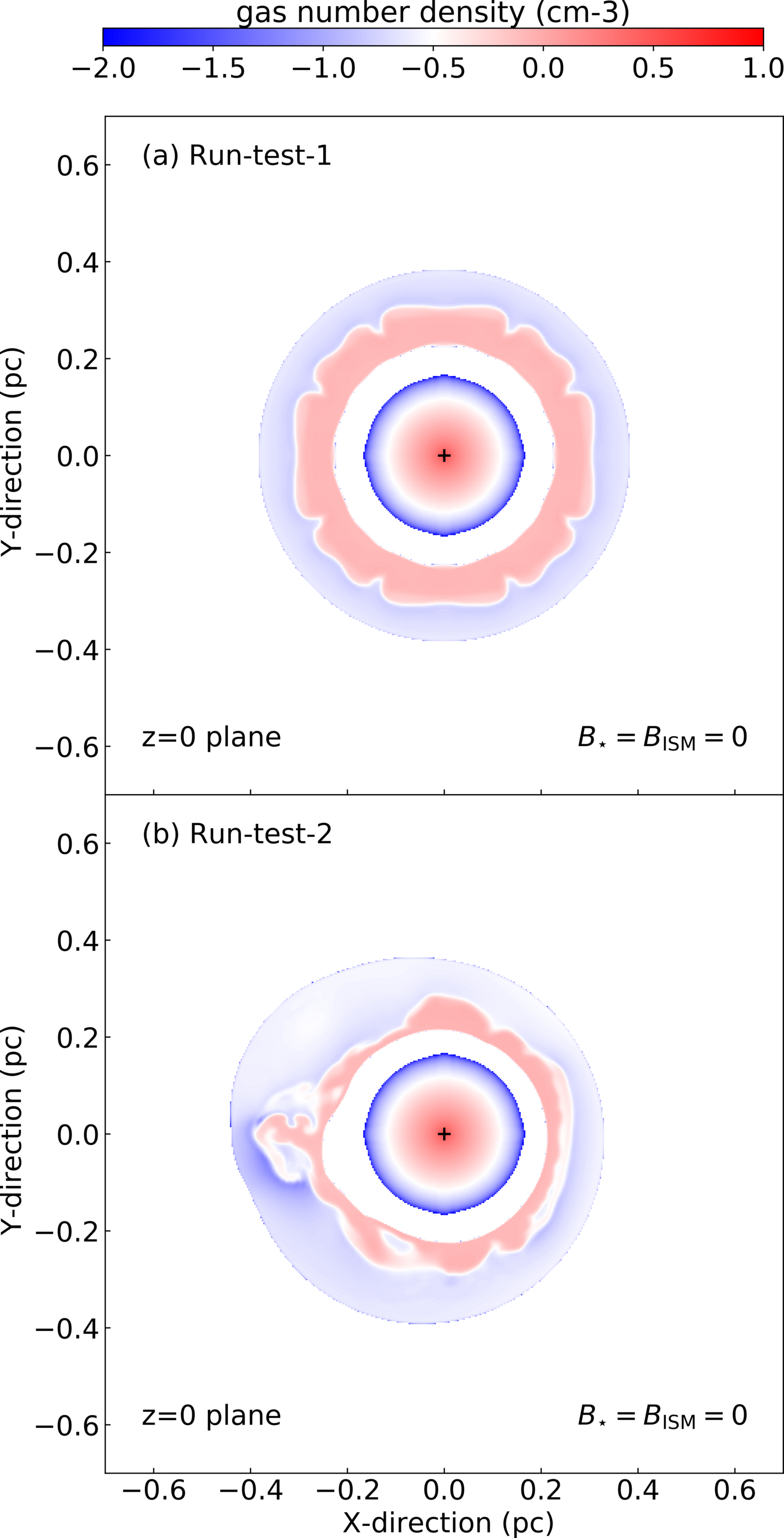}
        \caption{
        Density fields in the $z=0$ plane of models Run-test-1 and Run-test-2, 
        with $B_{\star}=B_{\rm ISM}=0\, \rm G$. 
        Difference between the two simulations is the direction along which 
        the star moves, i.e. along $Oz$ (a) and along an arbitrary direction 
        slightly deviating from $Oz$ (b). When the star runs parallel to 
        the vertical direction, instabilities develop symmetrically in the 
        bow shock, see also~\citet{blondin_na_57_1998}. 
        The black cross marks the position of the star. 
        }
        \label{fig:comparison}  
\end{figure}

\begin{figure}
        \centering
        \includegraphics[width=0.5\textwidth]{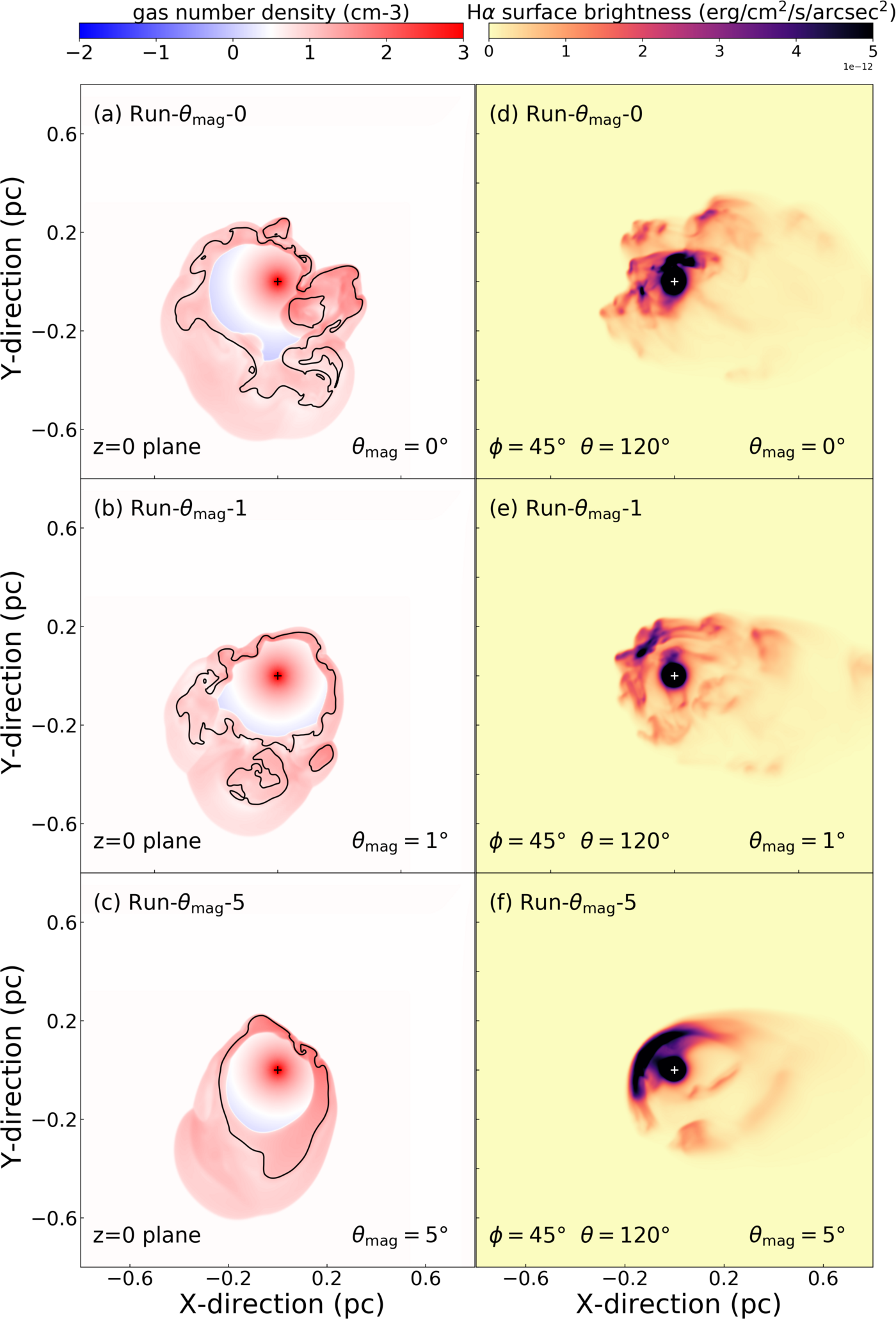}
        \caption{
        Density fields in the $z=0$ plane of models Run-$\theta_{\rm mag}$-0 and 
        Run-$\theta_{\rm mag}$-1 and Run-$\theta_{\rm mag}$-5, assuming different 
        angle between the direction of motion of the star and that of the local 
        magnetic field $\theta_{\rm mag}=0\degree$ (a), $\theta_{\rm mag}=1\degree$ 
        (b) and $\theta_{\rm mag}=5\degree$ (c), respectively. 
        The right part of the figure plots the corresponding surface brightness 
        (in $\rm erg\, \rm cm^{2}\, \rm s^{-1}\, \rm arcsec^{-2}$) assuming a 
        viewing angle $\phi=45\degree$ and $\theta=180\degree$, respectively. 
        The black contour is the location of the bow shock made of equal 
        proportion of stellar wind and ISM material. 
        The black (left) and white (right) crosses mark the position of the star. 
        }
        \label{fig:density_horizontal}  
\end{figure}

Previous 3D Eulerian numerical works on stellar wind bow shock showed that 
instabilities develop along the $Ox$ and $Oy$ axis when the star moves 
along the $Oz$ axis~\citep{blondin_na_57_1998}. 
In order to avoid such an effect, our simulations  impose the star a 
trajectory which does no coincide with a preferential direction of 
the 3D Cartesian coordinate system ($Ox$, $Oy$ and $Oz$). Our direction 
of stellar motion makes arbitrary angles $\leq 10\degree$ with 
the planes $Oxy$ and $Oxz$, respectively. 
In Fig.~\ref{fig:comparison}, we plot a series of two models labelled 
Run-test-1 and Run-test-2 (Tab.~\ref{tab:models}), which compare the 
above discussed grid effects. We carry out these tests within the 
hydrodynamical limit ($B_{\star}=B_{\rm ISM}=0$). The difference lies 
in the direction of the stellar motion imposed by the outer 
boundary conditions at the planes $x=0.8$, $y=0.8$ and $z=0.4$, 
respectively. In Run-test-1 the star moves along the $Oz$ axis 
while in Run-test-2 it moves along an arbitrary direction as 
described above. 

The most important differences are the changes in the topology of the 
different discontinuities present in these cross-sections. The 
termination shock, \textcolor{black}{astropause} interface and forward 
shock are spherically symmetric in Run-test-1 (Fig.~\ref{fig:comparison}a) 
while absent in Run-test-2 (Fig.~\ref{fig:comparison}b). This fact reflects 
different faster-growing modes at work in the two simulation tests, it 
confirms the findings of~\citet{blondin_na_57_1998} and it illustrates 
the necessity to impose the star a direction of motion which does not 
coincide with any Cartesian axis unless grid-induced regular patterns 
such as the $m=4$ mode of Fig.~\ref{fig:comparison}a will develop in 
the shocked stellar wind. Although both are physical, Run-test-2 is 
closer to what would happen in nature. 
%
%
The isothermal stellar wind bow shock of~\citet{blondin_na_57_1998} concerns 
a slow-, dense-winded star, moving with velocity $\approx 60\, \rm km\, \rm s^{-1}$ 
that develops a ragged, unstable, clumpy circumstellar structure. Their fig.~9 and 
10 show the bow shock structure strongly influenced by the grid during 
many crossing-time of the ISM flow through the computational domain before becoming 
turbulent. \textcolor{black}{Our method permits us to eliminate these grid effects without 
waiting for many flow times and reduce computational costs.}

\subsection{Comparison with observations}
\label{sect:discussion_irc10414}

\subsubsection{The case of IRC-10414}

\begin{figure*}
        \centering
        \includegraphics[width=0.95\textwidth]{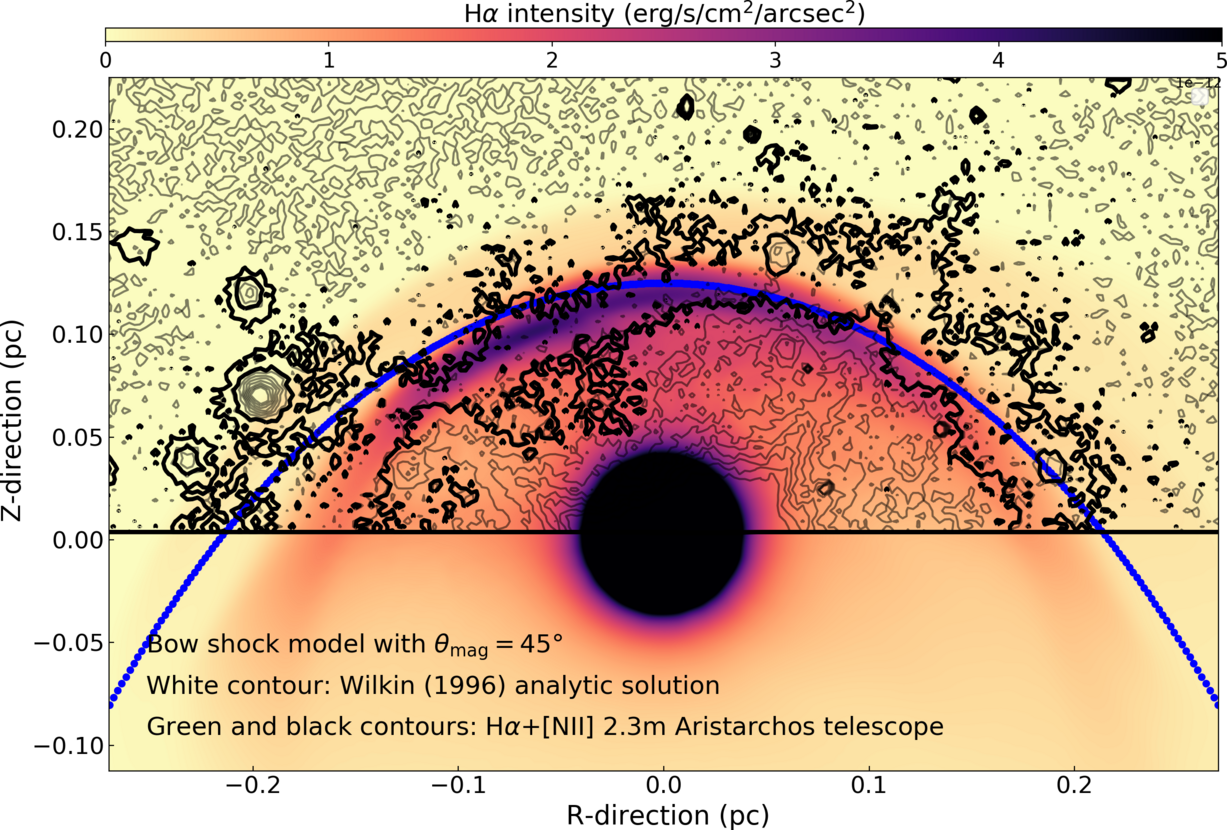}
        \caption{
        Comparison between our bow shock model Run-$\theta_\mathrm{mag}$-45 and 
        observational data of the circumstellar medium of IRC-10414. 
        The image plots the surface H$\alpha$ brightness 
        calculated from our model with $\theta_\mathrm{mag}=45\degree$ 
        (in $\rm erg\, \rm cm^{2}\, \rm s^{-1}\, \rm arcsec^{-2}$) and the 
        H$\alpha$+[NII] data taken by the 2.3 meter Aristarchos telescope, 
        as black contours. Thin contours are the background data and thick 
        contours trace the bow shock nebula. 
        The blue contour represents the analytic solution of~\citet{wilkin_459_apj_1996}. 
        }
        \label{fig:observations_1}  
\end{figure*}

For the sake of completeness, we perform a comparison with observations 
between our modelled bow shock, which stellar wind properties are 
tailored to that of IRC-10414, and observational data of that particular 
circumstellar nebula. 
In Fig.~\ref{fig:observations_1}, we plot a synthetic image of the simulation 
model Run-$\theta_{\rm mag}$-45 generating a stable nebula, as seen at the 
optical H$\alpha$ emission line with a viewing angle characterised by 
$\theta=90\degree$ and $\phi=120\degree$ and which emission maps 
has been rotated so that the symmetry axis of the bow shock fits 
with the vertical axis of the plot. 
The figure additionally shows several contours from the observations 
taken by the Aristarchos telescope from the National Observatory of Athene 
(NOA) located at Mt Helmos, Greece (2326 meters above sea level). 
The data have been acquired via a single 30 minutes exposure, 
on August 9th, 2013. The thin black lines are several H$\alpha$+[NII] 
surface brightness isocontours highlighting the image background sky, 
and two thick black isocontours trace the arced circumstellar nebulae of 
IRC-10414. Last, the blue contour marks the analytic solution derived 
by~\citet{wilkin_459_apj_1996}. \textcolor{black}{The data do not extend 
under $z=0$ as the telescope aperture was chosen to screen part of the star.}

The observed bow shock is stable in the sense that no large-scale instabilities  
have developed. However, overplotted isocontours indicate that the flux is 
in some region of the nebula somewhat diffused. Perhaps this is caused 
by contamination from the starlight or  background field sources, e.g. 
in the $R<0$ and $z>0$ part of the image. 
Despite of good qualitative agreements between our predictions and 
observations, the curvature of the bow shock is not the same 
in the $R<0$ and $R>0$ regions of the figure. This difference may rely on 
projection effects that our models do not reproduce, and better 
fine--tuning of our simulations  would be necessary to address 
this question \textcolor{black}{beyond the scope of our study, 
see~\citet{baalmann_aa_650_2021} for a 
first attempt. 
Therefore, even more,} complex 3D MHD models are necessary to fit the data 
better, e.g. considering local variations in the ambient 
medium density and/or the turbulent character of the medium the 
star moves in. We chose to fit the data by matching the region at 
the left part ahead of the apex of the bow shock. However, we could 
have chosen to further rotate the nebula and fit one of our radiative 
transfer model with the region at the right of the apex of the bow shock,  
which would have engendered simulation/observations discrepancies in the 
left-hand part of the structure. This argument is strengthened by the blue 
contour representing Wilkin's solution and consistent with only a part of 
the observed bow shock. 
Overall, our relatively good comparison with observations stresses 
that the bow shock of IRC-10414 is probably in a steady state. 
In other words, the star might have probably entered the red supergiant 
phase more than $\sim 0.1\, \rm Myr$ ago, which corresponds to the 
crossing-time of the simulation.

\begin{figure*}
        \centering
        \includegraphics[width=0.8\textwidth]{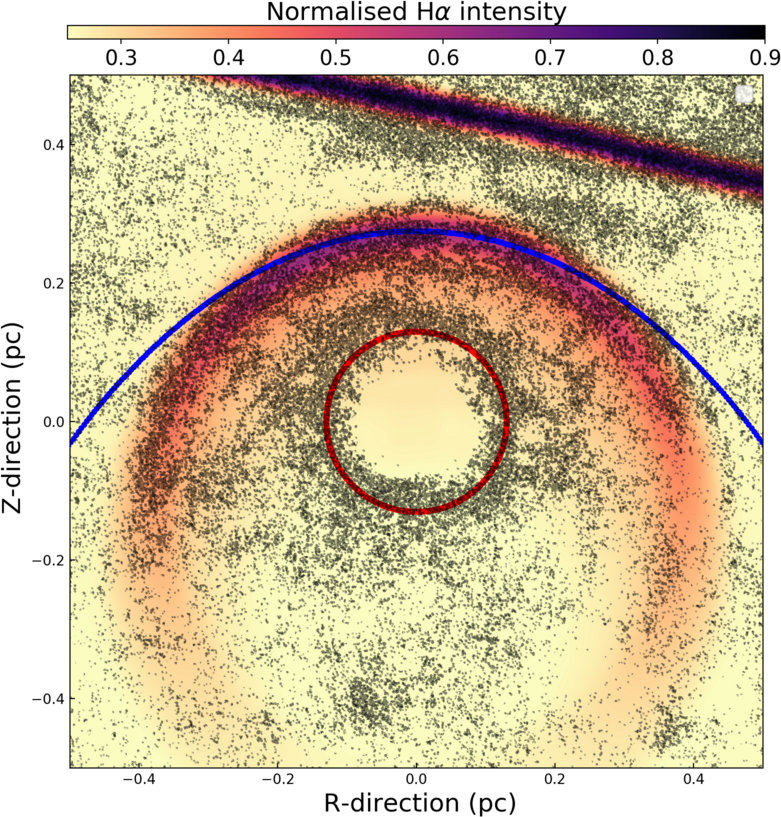}
        \caption{
        Comparison between our bow shock model Run-$\alpha$Ori and 
        observational data of the circumstellar medium of Betelgeuse. 
        The image plots the normalized H$\alpha$ surface brightness 
        calculated from our model Run-$\alpha$Ori tailored to Betelgeuse, 
        assuming with $\theta_\mathrm{mag}=90\degree$ 
        (in $\rm erg\, \rm cm^{2}\, \rm s^{-1}\, \rm arcsec^{-2}$) and the 
        infrared $170\, \mu \rm m$ taken by the space-borne telescope {\sc Herschel}  
        as black contours. The upper structure is the bar accompanying Betelgeuse's bow 
        shock~\citet{noriegacrespo_aj_114_1997}. 
        The blue line represents the analytic solution of~\citet{wilkin_459_apj_1996}, 
        the red circle marks the photo-ionizing confined shell interpreted and modelled 
        in~\citet{2014Natur.512..282M}. 
        }
        \label{fig:observations_2}  
\end{figure*}

\subsubsection{The case of Betelgeuse}

Betelgeuse ($\alpha$Orionis) is the Earth-closet red supergiant star located 
in \textcolor{black}{Orion's} neighbouring massive star-forming region. It is a runaway stellar 
object which displays a parsec-scale bow shock~\citep{noriegacrespo_aj_114_1997}. 
This object is a laboratory to study the physics of cool stellar winds such as dust 
and test stellar pulsations models of these variable stars. 
\textcolor{black}{
Betelgeuse is the priority target of active researches on photometric variability of 
cool stars such as red supergiant and asymptotic giant branch stars. These works 
permitted the monitoring of Betelgeuse's lightcurve and its Great optical 
Dimming~\citep{dupree_apj_899_2020,levesque_natur_594_2021} together with measures 
of its magnetic field~\citep{mathias_aa_615_2018,montarges_natur_594_2021}. 
}
Its intriguing circumstellar medium constituted of a stellar wind bow shock, an bright 
inner ring and a more extended, unexplained bar makes it an interesting target for
hydrodynamical modelling. 
Previous studies on the surrounding of Betelgeuse taught us that it is a 
runaway star which probably experienced a blue loop. It was recently returned 
to the red supergiant phase~\citep{mackey_apjlett_751_2012}, leaving behind its 
precedent blue supergiant bow shock after briefly forming Napoleon's 
hat~\citep{wang_MNRAS_261_1993}. 
The youngness of its stellar wind bow shock is also the main conclusion of the 
work of~\citet{mohamed_aa_541_2012}, presenting 3D hydrodynamical Lagrangian 
simulations tailored to the surroundings of $\alpha$Orionis. 
The smooth appearance of the astrosphere of Betelgeuse contradicting 
some simulation model of\citet{mohamed_aa_541_2012}. The inhibition of Rayleight-Taylor 
instabilities growing at the dense \textcolor{black}{tangential discontinuity} 
interface has been studied with 
2.5D MHD simulations in \citet{vanmarle_aa_561_2014}, showing that the 
ambient magnetic field of Orion's arm is sufficiently strong to reduce and damp 
the instabilities in the bow shock around Betelgeuse and modify its appearance. 
As for the astrosphere of IRC-10414, Betelgeuse is externally photoionised. 
The radiation pressure from the ionized ISM confines each of the \hii region 
of Betelgeuse, where the cool wind meets its hot accelerating wind within a 
neutral shell, is therefore located inside of the termination shock of 
the bow shock~\citep{2014Natur.512..282M}.

To further validate our method, we perform an additional 3D MHD simulation of 
Betelgeuse's astrosphere by adopting the parameters of~\citet{vanmarle_aa_561_2014}. 
In our simulation Run-$\alpha$Ori, the mass-loss rate of the star is 
$\dot{M}\approx3\times 10^{-6}\, \rm M_{\odot}\, \rm yr^{-1}$, the stellar 
wind velocity is $v_{\rm w}=15\, \rm km\, \rm s^{-1}$ and its space velocity 
$v_{\star}=28.3\ \rm km\, \rm s^{-1}$, respectively. The ISM ambient medium 
has a density $n_{\rm ISM}=1.89\, \rm cm^{-3}$, and we consider, as for IRC-10414, 
that its circumstellar medium is externally-ionized. Apart from the stellar wind and ISM 
properties, the numerical setup is the same as for IRC-10414. We adopt 
$\theta_{\rm mag}=90\degree$, i.e. we assume that the ambient medium 
magnetic field lines are nearly parallel to Betelgeuse's bar. 
%
%
The surroundings of Betelgeuse is more complex than that of IRC-10414. \textcolor{black}{Hence, we} 
proceed differently as in Fig.~\ref{fig:observations_1}. \textcolor{black}{ Betelgeuse's bowshock is not visible in H$\alpha$ because the star is too bright and its young nebulae too small;} we mask the material that is inside of the 
photoionised-confined shell (also not modelled) extending up to radius $\sim \, \rm pc$.
The bar is added as a denser region of Gaussian density profile peaking at 
$70\, \rm cm^{-3}$ and inclined with the direction of stellar motion with an 
angle that we adjust so that projected emission fits infrared observations. 
Fig.~\ref{fig:observations_2} compares an H$\alpha$ emission map of the 
bow shock of Betelgeuse and its bar, with overlaid {\sc PACS} 
{\sc Herschel}\footnote{ ESA Herschel Science Archive Observation ID: 1342242656.  } 
infrared data (black contours). The blue contour is Wilkin's solution for the 
bow shock morphology calculated with the observed \textcolor{black}{stand-off distance 
of Betelgeuse's astrosphere}, whereas the red circle marks the location of 
the neutral shell forming at the neutral-ionized wind interface~\citep{2014Natur.512..282M}. 
The projected bow shock structure matches well the infrared emission and we conclude 
that our simulation model for Betelgeuse as a $\simeq 15\, \rm kyr$ old red supergiant star 
which astrosphere is magnetically stabilized is a probable explanation for the 
circumstellar medium of that star. We do not need any stellar argument to explain 
the bar that we treat as independent from the \textcolor{black}{astropause} interaction region, 
without any stellar evolution argument at the origin of the formation of, e.g. 
a double bow shock structure of~\citet{mackey_apjlett_751_2012}. The concavity of 
the bar, slightly opposite of that of the 
astrosphere~\citep[see][]{noriegacrespo_aj_114_1997,vanmarle_aa_561_2014}, 
leads us to interpret the bar as of circumstellar origin, which supports 
the conclusions of~\citet{decin_2012}. The bow shock and \textcolor{black}{Betelgeuse's bar 
result} on a serendipitous disposition and projection effects from 
both components.

\begin{figure}
        \centering
        \includegraphics[width=0.45\textwidth]{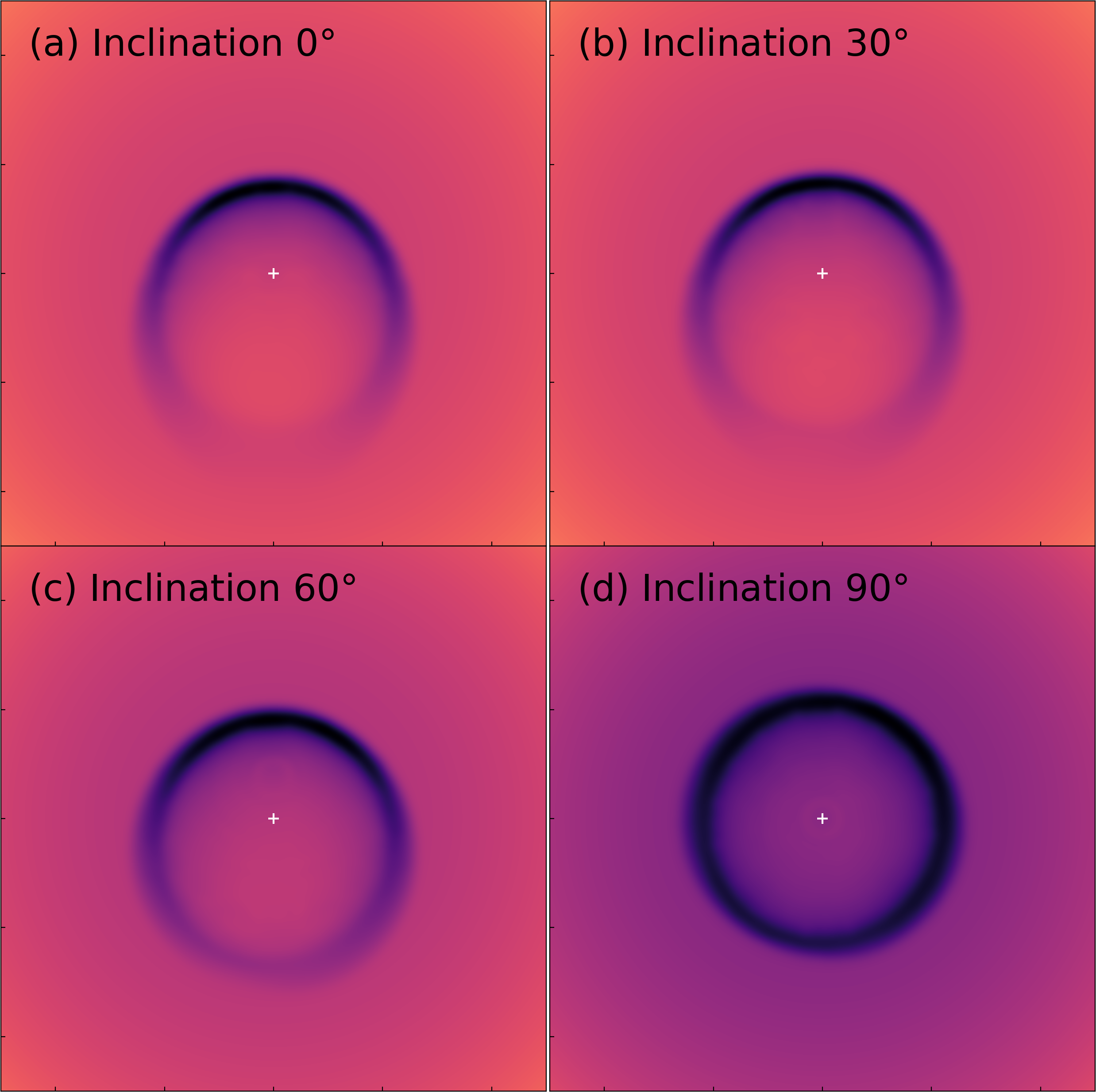}
        \caption{
        \textcolor{black}{
        Series of rendered emission maps highlighting the 
        effects of the inclination angles with respect to the plan of the sky, 
        for our best fit model from Run-$\alpha$Ori to Betelgeuse's bow shock. 
        We refer the reader to fig.~12 of~\citet{mohamed_aa_541_2012} for 
        a similar exercise.  
        The images are non-background subtracted and convolved with a 
        Gaussian filter.  
        The position of the star is marked by a white cross. 
        }
        }
        \label{fig:inclination_effects}  
\end{figure}

\textcolor{black}{
Last, this radiative transfer modelling for the surroundings of Betelgeuse calls a few 
comments. First, concerning the selected simulated snapshot that we decided to produce 
Fig.~\ref{fig:observations_2}. We took the youngest bow shock which aspect 
ratio is consistent with Betelgeuse's observed one, measured to $R(0)=0.275\, \rm pc$ 
and $R(0)/R(90) \approx 0.76$~\citep{mohamed_aa_541_2012,vanmarle_aa_561_2014}. 
It consists to the time instance $0.015\, \rm Myr$ after the onset of Run-$\alpha$Ori, 
when the physical size of the astropshere corresponds to that derived 
in~\citet{noriegacrespo_aj_114_1997}. 
The aspect ratio $R(0)/R(90)$ persists up to $0.4\, \rm Myr$. At the same time, the 
nebula grows, i.e. bow shocks at a later time will still reasonably match the {\sc Herschel} 
observations, however, at the cost of reevaluating the distance to the star. 
Secondly, the adopted inclination angle of the Betelgeuse bow shock with respect to the 
plane of the sky. For the sake of simplicity, when generating the initial conditions for 
the radiative transfer calculation, we have assumed an inclination angle of $0\degree$. Further 
modelling with the {\sc radmc-3d} code indicates that no noticeable differences appear for 
inclination angles $\le 30\degree$. However, for angles $> 45\degree$ the H$\alpha$ bow shock 
experiences significant projection effects, as already noted using column densities 
by~\citet{mohamed_aa_541_2012}, see also our Fig.~\ref{fig:inclination_effects}. 
}

\subsection{Implication for other astrophysical objects}
\label{sect:discussion_sn}

\subsubsection{Implication for bow shocks around OB stars}

The many projection effects at work in our optical H$\alpha$ emission maps of bow 
shocks around runaway red supergiant stars (Figs.~\ref{fig:maps_0}-\ref{fig:maps_90}) 
can be discussed in the context of earlier-type massive stars. 
Various catalogues of arc-like nebulae have been established, principally in 
the infrared waveband of the electromagnetic spectrum~\citep{vanburen_apj_329_1988,
vanburen_aj_110_1995,noriegacrespo_aj_113_1997,peri_aa_538_2012,peri_aa_578_2015,
kobulnicky_apj_710_2010,kobulnicky_apjs_227_2016,kobulnicky_aj_154_2017}.
Many \textcolor{black}{astrospheres} around those main-sequence massive OB stars 
display partly irregular shapes, making their detailed classification difficult, 
redundant and uneasy. At the light of the emission maps of our 3D MHD simulations, 
we propose that other bow shocks are affected by similar projection effects, 
and that the peculiar morphology of some bow shocks around earlier-type 
massive stars are hence produced. 
This would greatly reduce the taxonomy of different physical morphologies of 
astrospheres.

\subsubsection{Implication for core-collapse supernova remnants}

Besides constraining stellar evolutionary parameters, the prime importance 
for the study of the surroundings of evolved massive stars is their role 
as the pre-supernova circumstellar medium of core-collapse supernova 
progenitors~\citep{borkowski_apj_400_1992,velazquez_apj_649_2006,
toledo_mnras_442_2014,broersen_mnras_441_2014,2019arXiv190908947C,chiotellis_mnras_502_2021}. 
After the explosive death of a massive star, the supernova shock wave will 
inevitably interact with its pre-shaped surroundings, which, according to the 
mass that is trapped into it~\citep{meyer_mnras_450_2015}, first potentially 
feel its presence from the dynamical point of view, generate an asymmetric 
supernova remnant and eventually further expand into the ambient 
medium~\citep{meyer_mnras_493_2020}. 
Inversely, the morphology of core-collapse supernovae remnants permit to 
constrain the properties of its circumstellar medium, itself function of 
the past evolution of the progenitor star. 
A significant fraction of massive stars being runaway stellar objects, a similar 
fraction of core-collapse supernova remnants must have been shaped, or at least 
must have at some time beard imprints of their progenitor's bow shock, driven 
either by a red supergiant or a Wolf-Rayet star~\citep{katsuda_apj_863_2018}. 
Therefore, the expanding supernova shock wave properties will be completely 
different once it has passed through its circumstellar medium, and this will 
affect, e.g. particle acceleration mechanisms therein. Carefully understanding 
cosmic-ray acceleration in core-collapse supernovae requires an 
accurate treatment of its ambient medium prior to the explosion, and our 
work shows that this should be performed by means of 3D MHD simulations.


\section{Conclusion}
\label{sect:conclusion}

In this study, we explore the morphology and emission properties of 
the astrosphere generated around a runaway evolved, cool massive star. 
We present the first 3D MHD simulations of bow shocks forming in the vicinity 
of a runaway supergiant star, tailoring our simulations devoted to the surroundings 
of the M-type star IRC-10414~\citep{Gvaramadze_2013,meyer_2014a}. 
%
%
Our numerical simulations are 3D Cartesian models performed with the 
{\sc pluto} code~\citep{mignone_apj_170_2007,migmone_apjs_198_2012,vaidya_apj_865_2018},  
a well-tested tool for modelling \textcolor{black}{astropause} 
interactions~\citep{meyer_mnras_464_2017}. 
Constant adopted stellar wind properties are that of IRC-10414, constrained 
in~\citet{Gvaramadze_2013} and~\citet{meyer_2014a}, and we assume 
that the ISM in which the star moves is supported by a typical magnetic 
field of strength $7\, \mu \rm G$. Our free parameter is the orientation 
of the ISM magnetic field with respect to the direction of stellar motion. 
We run a simulation for $0.4\, \rm Myr$, which corresponds to a few 
dynamical crossing-time of the gas throughout the computational 
domain. For the sake of completeness, stellar rotation and magnetic 
field are included into the numerical setup as a Parker spiral. 
This 3D MHD study, therefore, permits exploring the effects of a 
non-aligned magnetic field on the structure and emission properties 
of the resulting stellar wind bow shock.

Magnetic fields stabilise astrospheres~\citep{vanmarle_aa_561_2014,
meyer_mnras_464_2017,scherer_mnras_493_2020}. 
Mainly, when the local ISM magnetic field is aligned with less than a 
few degrees with that of the stellar motion, the bow shock is ragged, 
clumpy and strongly unstable, while previous 2D simulations produced 
stable structures~\citep{meyer_2014a}. 
Axisymmetric patterns are broken, principally because of large scale 3D eddies developing 
at the \textcolor{black}{astropause} interface and affecting the overall morphology of the bow 
shock wings. Grid-induced effects are at work when the star moves along 
a characteristic axis of the 3D Cartesian coordinate system, while they 
vanish when the stellar motion is directed along an arbitrary direction. 
Such results stress the need for 3D MHD simulations in further tackling the problem 
of the circumstellar medium of massive stars~\citep{vanmarle_2015}.  
We find that the stellar magnetic field is dynamically unimportant in 
shaping the astrosphere of evolved,  red supergiant, massive stars, which 
is consistent with other studies devoted to, e.g., the 
heliosphere~\citep{pogorelov_apj_845_2017} \textcolor{black}{and the study on 
$\lambda$ Cephei of~\citet{scherer_mnras_493_2020}. }

Synthetic optical H$\alpha$ emission maps of 3D MHD \textcolor{black}{astrospheres} 
around cool stars show that projection effects are important in the observed shape of the nebulae, 
and we speculate that this must also affect the astrospheres of earlier-type massive stars.
%
%
This study shows that the problem of the smooth appearance of red supergiant 
bow shock can be solved with a simple, mild magnetisation of its ambient 
medium, and that the previously proposed external ionisation and/or youth 
of the structure might not be enough to explain the \textcolor{black}{unique} appearance 
of both IRC-10414 and Betelgeuse, although it participates to 
it~\citep{vanmarle_aa_561_2014}. We conclude that IRC-10414 is a star 
in a steady state and that Betelgeuse's bar is of interstellar origin. 
\textcolor{black}{
Our work is therefore in accordance with the bow shock models for Betelgeuse 
of~\citet{mohamed_aa_541_2012} and~\citet{vanmarle_aa_561_2014}, but it is in 
clear disfavour regarding to that proposed in~\citet{mackey_apjlett_751_2012}. 
}
Our results will serve for future modelling of the circumstellar 
medium of runaway red supergiant stars and to realistically simulate the 
early supernova-blastwave interaction happening in core-collapse supernova 
remnants.

%
%
%


\section*{Acknowledgements}

D.~M.-A. Meyer thanks L.~Decin and A.-J.~van Marle for advice on observational data. 
The authors acknowledge the North-German Supercomputing Alliance (HLRN) for providing 
HPC resources that have contributed to the research results reported in this paper. 
M.~Petrov acknowledges the Max Planck Computing and Data Facility (MPCDF) for 
providing data storage resources and HPC resources which contributed to test and 
optimise the {\sc pluto} code. 
PFV acknowledges the financial support for PAPIIT-UNAM grant IA103121. 
This work is based on observations made with
the “Aristarchos” telescope operated on the Helmos Observatory
by the Institute of Astronomy, Astrophysics, Space Applications
and Remote Sensing of the National Observatory of Athens.

\section*{Data availability}

This research made use of the {\sc pluto} code developed at the University of Torino  
by A.~Mignone and collaborators (\url{http://plutocode.ph.unito.it/}) and of the 
{\sc radmc-3d} code developed by C.~Dullemond and collaborators at the University of Heidelberg 
(\url{https://www.ita.uni-heidelberg.de/~dullemond/software/radmc-3d/}), respectively. 
The figures have been produced using the Matplotlib plotting 
library for the Python programming language (\url{https://matplotlib.org/}). 
The 3D renderings have been generated using the Visit software 
{\url{https://visit-dav.github.io/visit-website/}}. 
Observational data of the circumstellar medium of Betelgeuse have been obtained 
via the Herschel data Schearch of the NASA/IPAC Infrared Science Archive  
(\url{https://irsa.ipac.caltech.edu/applications/Herschel/}).
The data underlying this article will be shared on reasonable request to 
the corresponding author.


\bibliographystyle{mn2e}

\footnotesize{
\bibliography{grid}
}

\end{document}